\documentclass{article}
\usepackage{arxiv}

\usepackage{amssymb}
\usepackage{lineno}

\usepackage{caption}
\usepackage{float}
\usepackage{mathtools}

\usepackage[utf8]{inputenc} 
\usepackage[T1]{fontenc}
\usepackage{natbib}
\setcitestyle{square} 
\usepackage[colorlinks=true, allcolors=blue]{hyperref}
\usepackage{eqparbox}
\usepackage{graphicx}
\usepackage{arydshln}
\usepackage[mathscr]{euscript}
\DeclareSymbolFont{rsfs}{U}{rsfs}{m}{n}
\DeclareSymbolFontAlphabet{\mathscrsfs}{rsfs}

\usepackage{authblk}
\title{Parameter sensitivity analysis of a sea ice melt pond parametrisation and its emulation using neural networks}
\author[1,*]{Simon Driscoll}
\author[2,1]{Alberto Carrassi}
\author[3]{Julien Brajard}
\author[3]{Laurent Bertino}
\author[4]{Marc Bocquet}
\author[3]{Einar {\"O}rn {\'O}lason}

\affil[1]{Department of Meteorology\\
	University of Reading\\
	United Kingdom, RG6 6BB}
\affil[2]{Department of Physics and Astronomy ``Augusto Righi''\\
	University of Bologna, Viale Carlo Berti Pichat, 6/2, Bologna, 40127}
 \affil[3]{Nansen Environmental and Remote Sensing Center, Jahnebakken 3, Bergen, N-5007, Norway}
 \affil[4]{CEREA, {\'E}cole des Ponts and EDF R\&D, \^Ile-de-France, France}
\affil[*]{\texttt{Corresponding author: s.driscoll@pgr.reading.ac.uk}} 

\date{11 April 2023}                     
\begin{document}
  \maketitle
\renewcommand{\headeright}{}
\renewcommand{\undertitle}{}
\begin{abstract}

Accurate simulation of sea ice is critical for predictions of future Arctic sea ice loss, looming climate change impacts, and more. A key feature in Arctic sea ice is the formation of melt ponds. Each year melt ponds develop on the surface of the ice and primarily via affecting the albedo, they have an enormous effect on the energy budget and climate of the Arctic. As melt ponds are subgrid scale and their evolution occurs due to a number of competing, poorly understood factors, their representation in models is parametrised.

Sobol sensitivity analysis, a form of variance based global sensitivity analysis is performed on an advanced melt pond parametrisation (MPP), in Icepack, a state-of-the-art thermodynamic column sea ice model. Results show that the model is very sensitive to changing its uncertain MPP parameter values, and that these have varying influences over model predictions both spatially and temporally. Such extreme sensitivity to parameters makes MPPs a potential source of prediction error in sea-ice model, given that the (often many) parameters in MPPs are usually poorly known. 

Machine learning (ML) techniques have shown great potential in learning and replacing subgrid scale processes in models. Given the complexity of melt pond physics and the need for accurate parameter values in MPPs, we propose an alternative data-driven MPPs that would prioritise the accuracy of albedo predictions. In particular, we constructed MPPs based either on linear regression or on nonlinear neural networks, and investigate if they could substitute the original physics-based MPP in Icepack. 

Our results shown that linear regression are insufficient as emulators, whilst neural networks can learn and emulate the MPP in Icepack very reliably. Icepack with the MPPs based on neural networks only slightly deviates from the original Icepack and overall offers the same long term model behaviour.

We then searched for the smallest possible emulator that achieves good performance by performing features selection based on mutual information. Results indicates that a smaller model, based only on a portion of the full set of input variables needed by the physical MPP, 
is also sufficient to approximate and replace the physical MPP. This smaller emulators are not only computationally faster but also easier to interpret on a physical ground. 

Several and diverse challenges still exist, yet this study is an encouraging first step, prior to using real data, toward the adoption of data-driven MPPs in sea-ice models. 


\end{abstract}

\section{Introduction}

Arctic sea ice plays an essential role in global ocean circulation and in regulating Earth's climate and weather \citep[e.g.,][]{Menemenlis2008,Sevellec2017,Dethloff2019}. Growing from a minimum in September, until the Arctic ice pack reaches its maximum extent in March \citep{Kwok_2018}, sea ice is a good thermal insulator, and also becomes a platform for an even greater insulator, snow. It acts as a lid on the surface of the polar oceans, and governs transfer of heat and water vapour between a comparatively warmer ocean and cold atmosphere \citep{Weeks2010}. 

Of fundamental importance in its own right, sea ice is also one of the Earth system components most sensitive to climate change \citep{Castellani2020}, and reductions in Arctic ice cover have been considerable over the satellite record since 1979 \citep{Serreze2007, Stroeve2012}. Climate models project that Arctic sea ice loss will continue with the possibility of ice-free summers occurring before or by the end of the 21st century \citep[e.g.,][]{Boe2009,Wang2009}. Losing high-albedo sea ice causes exposure of the dark ocean surface, more sunlight is absorbed, which in turn enhances surface warming \citep{Manabe1980}. This feedback plays a major role in the Arctic amplification of global warming \citep{Hall2004}. 


A key question for the climate system, with important biological \citep{Solan2020}, economic \citep{Alvarez2020} and geopolitical \citep{Huntington2022} implications, is understanding when sea ice may entirely vanish from Arctic summers. Climate models have traditionally significantly underestimated the observed trend in Arctic sea ice decline \citep{Boe2009}, and have also failed to capture physical changes in other warming periods, such as the elevated Arctic temperatures during the last interglacial (LIG) period \citep{Guarino2020}. 

\subsection{Sea-ice melt ponds and their role in Arctic climate}

Whilst reductions in sea ice and leads opening between the ice alter the albedo and thus the climate of the Arctic by exposing the ocean surface, another profound impact to the Arctic's climate occurs on the ice itself, through the formation of {\it melt ponds}. Melt ponds develop over the sea ice during the melting season from the accumulation of melt water from ice and snow, and may cover up to 50\% of the Arctic sea ice surface itself \citep{Flocco2015}. Whilst the albedo of bare sea ice can be up to $0.8$, melt ponds may have an albedo as low as $0.15$ \citep{Flocco2015}, and their evolution on the Arctic sea ice in summer is one of the main factors affecting sea ice albedo and hence the polar climate system \citep{Li2020}. 

The seasonal evolution of Arctic sea ice albedo undergoes five distinct phases: dry snow, melting snow, pond formation, pond evolution, and freeze-up during Autumn \citep{Perovich2002}. In April (with much of dry snow) the albedo tends to be high, and generally uniform, whilst as summer progresses it lowers and becomes spatially inhomogeneous. Accounting for these phases in numerical models of the sea ice requires a parametrisation of melt pond effects. Melt ponds shape, size, and coverage are determined by a number of factors that include ice surface topography and the total water amount available \citep{Holland2012}. Melt ponds exist at a higher percentage over thinner younger ice (at the end of the Artic summer) and the impact of melt ponds will increase, as the area of young ice is expected to increase due to climate change. The role and the impact of melt ponds on the physics and dynamics of the sea ice is also evidenced by numerical experiments with sea ice models. The latter exhibit a large sensitivity, particularly in terms of ice thickness to melt pond parametrisations \citep[e.g.,][]{Ebert1993}. \citet{Flocco2010,Flocco2012} have shown that models lacking a melt pond parametrisation (MPP) can overestimate the summertime sea ice thickness by up to 40\%. 


Other results indicate the importance of melt pond processes and point to the need for their accurate simulation. With improved model physics and a sophisticated melt pond parametrisation, \citet{Guarino2020} showed that the Hadley Centre HadGEM3 climate model hindcasted a complete loss of Arctic summer sea ice in LIG period simulations. They concluded that loss of Arctic sea ice in their simulations is largely due to increased net short-wave radiation, and not to a slowdown in the Atlantic Meridional Overturning Circulation (AMOC) as previously hypothesised. Their results suggested that local thermodynamic and melt pond formation processes play a key role in capturing the Arctic sea ice loss. Therefore, skilful simulations of future Arctic sea ice extension calls for accurate simulations of MPPs.  


\subsection{Using machine learning for parametrisation of physical processes in climate models}

Machine learning (ML) can accurately and efficiently recognise complex patterns and emulate nonlinear dynamics \citep{Kashinath2021}. Interest in using ML to emulate parametrised processes in climate simulations has grown considerably in the last decade \citep[e.g.,][]{Brenowitz2018,Rasp2018,Chantry2021}, largely (but not only) since emulators may offer improvements in computation speed as compared to traditional physics-based parametrisations \citep{Thiagarajan2020}. When trained against high resolution models that explicitly resolves the processes intended to be parametrised, ML-based parametrisations showed potential to improve model skills \citep[e.g.,][for the case of moist convection in global climate models]{Krasnopolsky2013a,Krasnopolsky2013b,OGorman2018}. This ``perfect model scenario'' whereby training is based on high-resolution models, is functional to the chronic lack of sufficient spatio-temporal dense accurate dataset required by proper training. This issue can be combated by combining data assimilation methods with ML, in such a way that the former handles efficiently the sparse and noise data and provides the analysed fields to train the ML \citep{Brajard2020,Brajard2021}.

\subsection{Sensitivity of melt ponds parametrisations and their discovery using machine learning}

This work has two connected goals. In the first part (Sect.~\ref{ssa_of_mpp}) we study in detail the parametric sensitivity of the Icepack model, a state-of-the-art model of the thermodynamics processes in the sea ice \citep{Hunke2021}, to its physics-based MPP. We adopt a rigorous approach to quantify this sensitivity, using the Sobol sensitivity analysis method \citep{Sobol1993, Sobol2001, Saltelli2004}. As mentioned earlier inappropriate MPPs in sea ice model can cause overestimate of the summertime sea ice thickness by up to 40\% \citep{Flocco2010,Flocco2012}. Understanding the sources of uncertainty in MPPs and reducing them are both key to accurate climate simulations.

We shall see in our study that Icepack shows an intricate, generally high, dependency on the parameters of its MPPs, leading to large impacts in the predictions of crucial sea ice quantities. Understanding the source of uncertainty, narrowing it, and fine tuning the parameters is paramount. However, this is usually difficult to achieve as the true physics and geometry of melt ponds are very complex, and thus the MPP are usually over-parametrised.

To bypass the undesirable sensitivity of the physical MPP, and the limited accuracy against the costs of tuning the parameters, in the second part of this study (Sect.~\ref{emulation_of_mpp}), we investigate the capability of modern ML to emulate the MPPs with the fewest possible inputs. We shall show that neural networks (NN) can successfully learn the functional relationship between inputs and outputs in the MPP, and that this ML-based MPP can substitute Icepack into the full sea-ice model providing accurate and stable simulations. Finally, in Sect.~\ref{fs_using_mi} we search for the smallest possible, and thus interpretable, ML-based MPP by performing feature selection based on mutual information tools.  

We work under the ``perfect model'' scenario, whereby the training relies on the output of the MPP intended to be emulated, thus a first step toward using real data that will be considered in the follow-up study. we will allude to this point in our conclusions and discussion on future directions in Sect.~\ref{disc_and_future}. 

\section{The sea ice model \emph{Icepack} and its melt pond parametrisations} \label{icepack_and_its_mpss}

\begin{figure}
\includegraphics[width=0.9\textwidth]{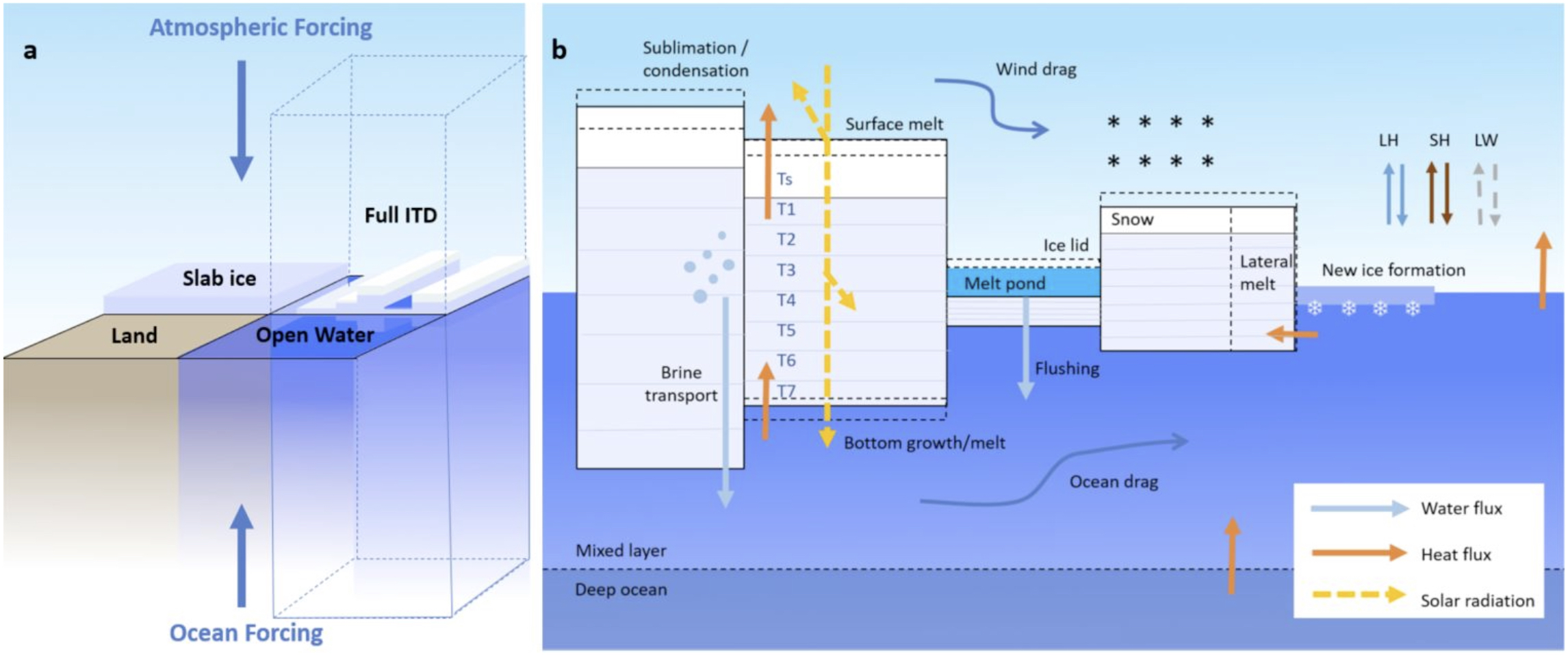}
    \caption{Schematic overview of the Icepack model. Panel a: the columnar model set-up, with 4 tiles representing land, open water, slab ice (without snow) and a full ice-thickness distribution (ITD). Panel b: the most important vertical processes represented in Icepack when using the ITD. Courtesy of \citet{Hofsteenge2020}. }
\label{fig:icepack_schematic}
\end{figure}

The Los Alamos Sea Ice Model (CICE) was developed to create an efficient sea ice component for a fully coupled atmosphere-ice-ocean-land global climate model \citep{Hunke2021CICE}. CICE has different interactive components: an ice dynamics model, a transport model, and the sub module {\it Icepack} that simulates all vertical processes, mainly of thermodynamics nature, in CICE \citep{Hunke2021}. Icepack module can be used as a stand-alone model and this is the way we shall implement it in this work. A schematic of Icepack together with the illustration of the important vertical processes are given in Fig.~\ref{fig:icepack_schematic}.  

\subsection{Melt pond parametrisations in Icepack}

Formation of melt ponds on ice is simulated in Icepack considering pond area and depth. When the top of the pond refreezes and snowfall on top of the refrozen lid blocks solar radiation, the ``effective pond area'' that is used for the radiation calculations can decrease while the pond volume remains constant. The effective pond area is therefore what influences the sea ice albedo. Three different schemes are available in Icepack to explicitly model melt ponds: (1) {\it CESM} \citep{Holland2012}, {\it topo} \citep{Flocco2010},  and {\it level-ice} \citep{Hunke2013}.

The first MPP is made specifically for the Community Earth System Model (CESM). In the CESM scheme, ponds can grow when rain or snow and ice melt water is added and shrink through refreezing. The melt pond processes are described empirically in this scheme and pond depth and area are linearly related. 

The {\it topo} scheme is said  topographic in that it simulates the concept that melt water collects on the lowest parts of the ice \citep{Flocco2007}. Since Icepack does not explicitly model ice topography, the ice thickness distribution is split into a surface height and basal depth distribution relative to sea level. Melt water is collected on the ice of the lowest surface height. In this scheme pond water can refreeze (affecting the effective pond area) and drain vertically when the sea ice becomes permeable. 

The last pond scheme, the {\it level-ice} \citep{Hunke2013}, builds upon {\it topo} but in additions it also accounts for gravity effects in a way where ponds can only form on the level (undeformed) ice areas per ice category. In the level-ice scheme, melt pond water can also refreeze and drain to the ocean, depending on the permeability of the ice. 

Being the most advanced formulation in Icepack \citep{Hunke2021}, in this study we focus on the level-ice MPP only, and shall hereafter refer to it as the MPP, without the need for further specification. 

\subsection{Icepack's melt pond parameters}

Icepack has a total of seven melt pond parameters \citep{Hunke2021}, although 
not all of them are used in the level-ice MPP. We know that one parameter is used only in the "topo" scheme, whilst the parameter \emph{dpscale}, the ``alter e-folding time scale for flushing pond refreezing parametrisation'', is not used  when the mushy thermodynamics option \citep[that treats the sea ice as a mushy layer in which a salinity profile is allowed to evolve, as in][]{Turner2013} is used. See \citet{Hunke2021} for more details.

Table~\ref{mp_params} lists all melt pond parameters in the level-ice MPP together with uncertainty ranges as given by \citet{Urrego-Blanco2016}. As can be seen, given these uncertainties, understanding the impact of these on the model is of crucial importance. It motivates us to employ a Sobol sensitivity analysis (SSA) on these parameters and explore the sensitivity of the sea ice variables in the Icepack model to these parameter values.

\begin{table}
\resizebox{\textwidth}{!}{%
\begin{tabular}{| c | c|c |c |c| c|} 
\hline
\multicolumn{6}{|c|}{\textbf{Parameters perturbed for SSA of the {\it level-ice} MPP in Icepack}} \\
\hline
parameter symbol & parameter description & prior probability distribution & min value & max value & default value \\ [0.5ex] 
\hline
hs0 & snow depth of transition to bare ice (m) & triangular & 0 & 0.1 & 0.03 \\ 
\hdashline
hs1 & snow depth of transition to pond ice (m) & triangular & 0 & 0.1 & 0.03 \\ 
\hdashline
rfracmin & minimum retained fraction of meltwater & triangular & 0 & 1 & 0.15 \\
\hdashline
rfracmax & maximum retained fraction of meltwater & triangular & 0 & 1 & 0.85 \\
\hdashline
pndaspect & ratio of pond depth to pond fraction & uniform & 0.4 & 1.2 & 0.8 \\ 
 \hline
\end{tabular}
}
\captionof{table}{\label{mp_params} Parameter symbols, descriptions and statistical distributions as given by \citet{Urrego-Blanco2016}. The distributions we use for SSA are visualised in Fig.~\ref{fig:kde_sobol}.}
\end{table}

\begin{figure}
\includegraphics[scale=0.21]{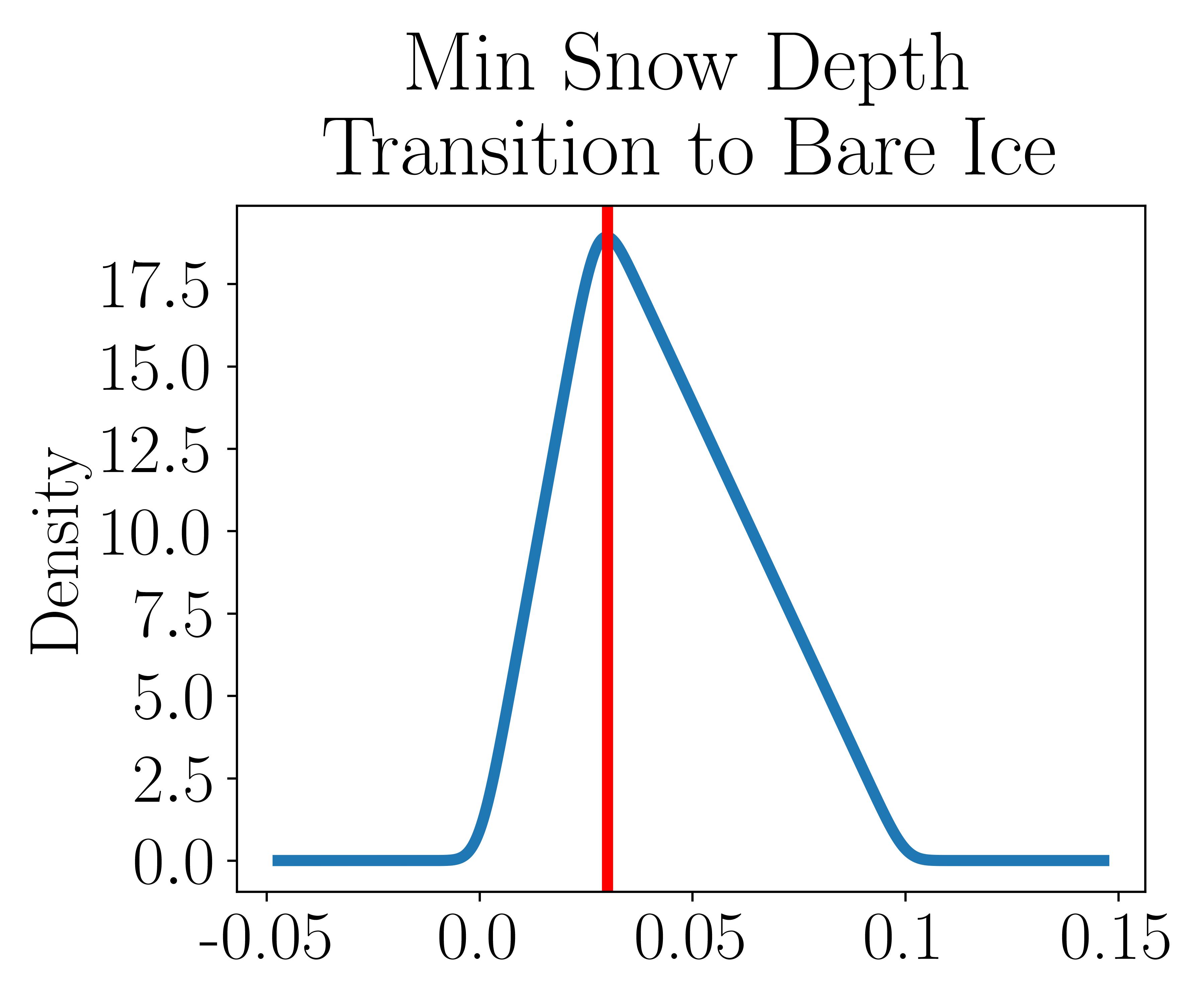}
\includegraphics[scale=0.21]{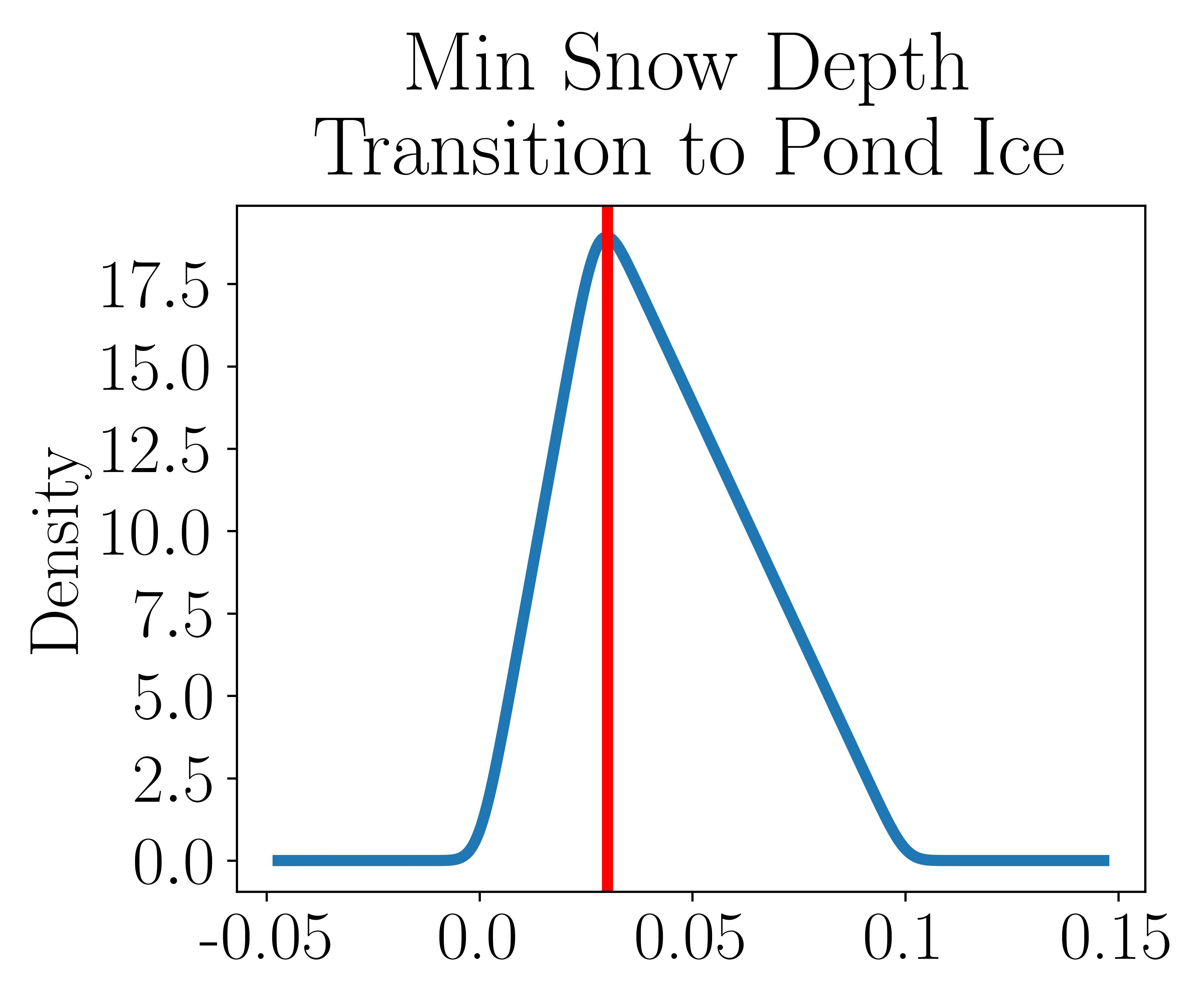}
\includegraphics[scale=0.21]{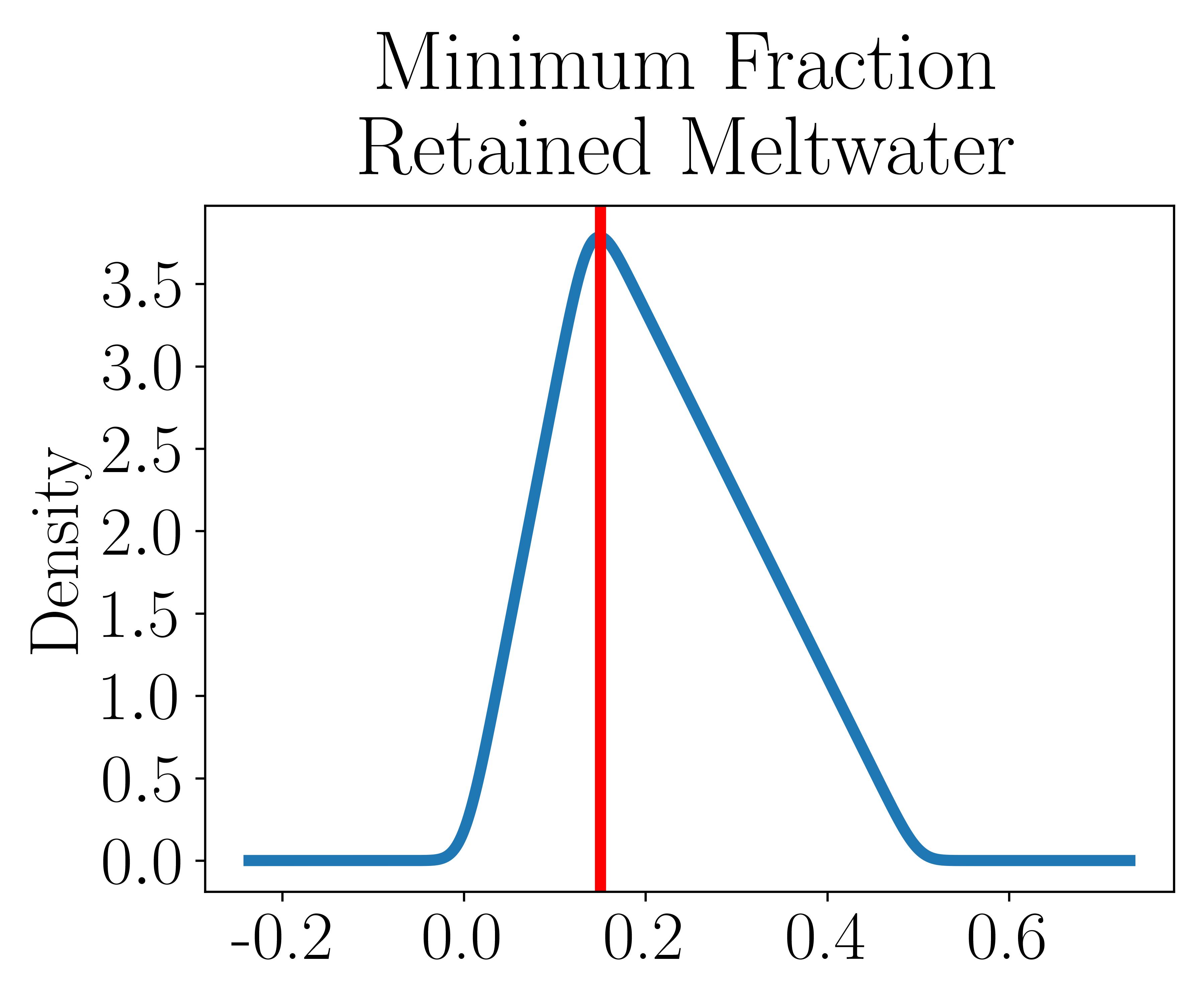}
\includegraphics[scale=0.21]{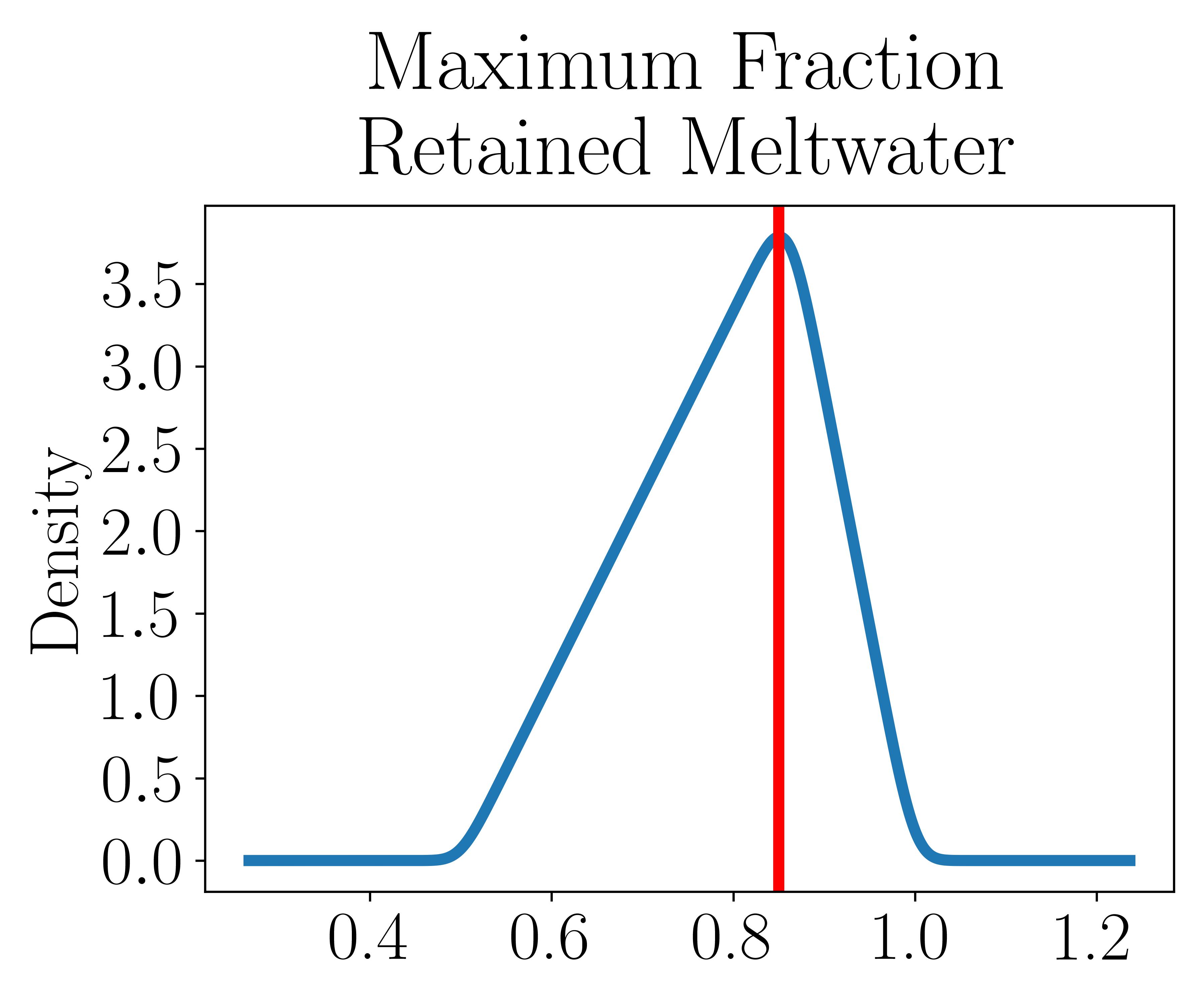}
\includegraphics[scale=0.21]{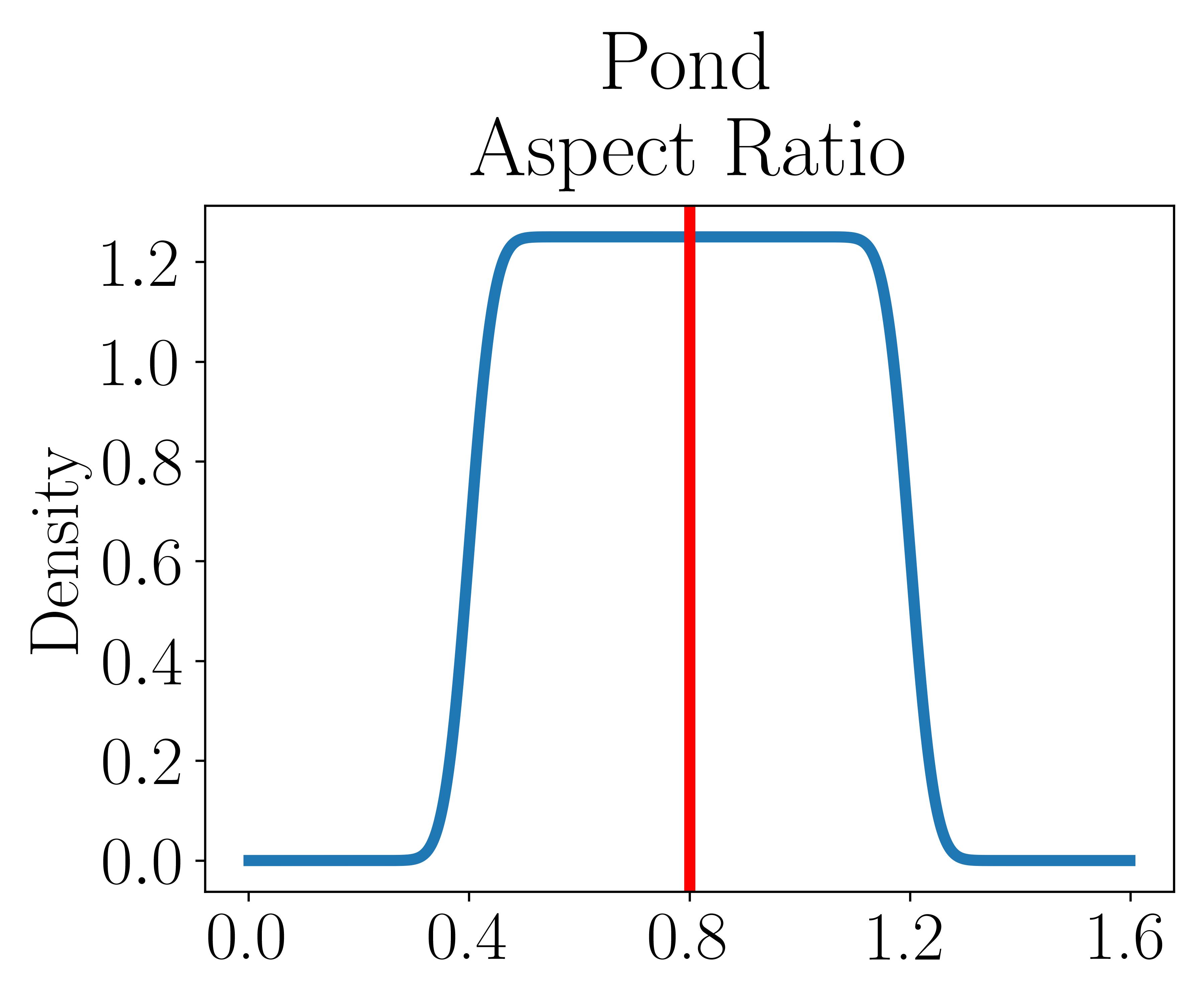}
\centering
\caption{Probability distribution functions we sample from using the SALib library for sensitivity analysis, for the 5 parameters invoked in the level-ice melt pond parametrisation with the mushy thermodynamics scheme (triangular and uniform). Red lines indicate default parameter values in the model. So that the minimum retained fraction of meltwater does not exceed the maximum, we bound the minimum to $[0,0.5]$ and the maximum to $[0.5,1]$.}
\label{fig:kde_sobol}
\end{figure}

\section{Sobol sensitivity analysis of melt pond parameters} \label{ssa_of_mpp}

\subsection{An outline of Sobol sensitivity analysis} 

Sensitivity analysis may be defined as ``the study of how uncertainty in the output of a model (numerical or otherwise) can be apportioned to different sources of uncertainty in the model input'' \citep{Saltelli2004}. Here we choose to perform a Sobol Sensitivity Analysis \citep[SSA,][]{Sobol1993, Sobol2001} on the Icepack MPP parameters. A global, variance-based method, SSA attributes variance in model output to individual or multiple parameters simultaneously, by taking into account the interactions between them \citep{Sobol2001,SALTELLI2002}. In studies of hydrological models, for instance, SSA provided sensitivity indices with best accuracy and robustness, especially with models with strong parameter interactions \citep[e.g.,][]{Tang2007}. 

For a detailed description and explanation of the SSA, see e.g., \citet{Sobol1993,Sobol2001,Saltelli2004}. We briefly outline it in the following. Let us consider the model:
\begin{equation}
    {\bf Y}= f(\boldsymbol{\lambda}) = f(\lambda_1, \lambda_2, ..., \lambda_d),
\end{equation}
with ${\bf Y}\in\mathbb{R}^m$ the model state vector, $\lambda_i\in\mathbb{R}$, for $i=1,...,d$, the model parameters, thought to be independent random variables, described by known distributions that reflect uncertain knowledge on the system. Let $\mathscrsfs{D}=\left\{1,...,d\right\}$. With $\Psi_i \subseteq \mathscrsfs{D}$ being the set containing the $i$-th parameter, we define $\tilde{\Psi}_i \coloneqq \mathscrsfs{D} \backslash i$, as its complement set in $\mathscrsfs{D}$, such that:
\begin{equation}
    \Psi_i \cup \tilde{\Psi}_i = \mathscrsfs{D},\qquad \Psi_i \cap \tilde{\Psi}_i = \emptyset.
\end{equation}
Assume that we know the true parameter values, where $\lambda_i=\lambda_i^*$, for $i=1,...,d$, and we would like to understand the impact of each parameter $\lambda_i$ upon the variance of the output ${\bf Y}$. We obtain the first order sensitivity index for parameter $\lambda_i$:
\begin{equation}
 {\rm S}_i=\frac{\mathbb{V}_{\Psi_i}(\mathbb{E}_{\tilde{\Psi}_i}({\bf Y}|\lambda_i))}{\mathbb{V}({\bf Y})},
\end{equation}
where $\mathbb{E}$ and $\mathbb{V}$ are the expectation and variance operator, respectively.

First order sensitivity indices characterise the fraction of the variance due to the parameter $\lambda_i$ only, i.e. without interaction with others, and from the normalised law of total variance, we see that ${\rm S}_i \leq 1$. 

Total sensitivity indices have been introduced in \citet{HOMMA1996}: they account for all the contributions to the variation in output due to parameter $\lambda_i$ plus all its interactions with the other parameters. Total indices can be given as 
\begin{equation}
     {\rm ST}_{i} = \frac{\mathbb{E}_{\tilde{\Psi}_i}(\mathbb{V}_{\lambda_i}({\bf Y}|\tilde{\Psi}_i))}{\mathbb{V}({\bf Y})},
\end{equation}
where ${\tilde{\Psi}_i}$ denotes all uncertain parameters except $\Psi_i$, and we have therefore $0 \leq {\rm S}_{i} \leq {\rm ST}_{i} \leq 1$, implying by definition that the total index will always bound the normal index from above. 

\subsection{SSA of Icepack parameters}

To conduct Sobol sensitivity analysis on the Icepack model we use the Sensitivity Analysis Library in Python \citep[SALib,][]{Herman2017}. Parameter uncertainty estimates come from \citet{Urrego-Blanco2016}, adapted to be suitable for SALib. 
Whilst the logit-normal probability distribution function as given by \citet{Urrego-Blanco2016} for certain parameters is not available in SALib, our aim is to investigate sensitivity to parameter values, rather than reducing parameter uncertainty, thus we approximate this with a triangular distribution. These input parameters (and their uncertainty range) for Icepack's melt pond parameters are given in Tab.~\ref{mp_params}. Using SALib, a global sample of this parameter space is taken using Sobol-Saltelli sampling to achieve a uniform coverage of the space \citep{Sobol2001,Saltelli2008}. 

Icepack is a 1D thermodynamic column model that runs at a single geographic location and it requires forcing data to provide outputs of sea ice area fraction, average ice thickness, and so on.  The data we choose to force Icepack with comes from the NCEP Climate Forecast System Version 2 (CFSv2) \citep{Saha2014}. CFSv2 is a fully coupled model representing the interaction between the Earth's atmosphere, oceans, land and sea ice. Initialized four times per day (00:00, 06:00, 12:00, and 18:00 UTC), it provides us with more than a decade (2011-2022 period) of hourly data over many Arctic locations with which to force Icepack. 

The Saltelli sampler of a provided parameter space in the SALib library generates $N\times(2d+2)$ samples, where $N$ is a chosen integer, required to be a power of $2$ for equal subsampling and $d$ is the number of parameter inputs. A low value of $N$ would imply insufficient sampling of the parameter space. In this study to guarantee a sufficient sampling of our parameter space, we use $N=128$, with $d=5$ parameters (cf Tab.\ref{mp_params}), yielding $1,536$ simulations at each location. Icepack is then run with these combinations of perturbed parameter inputs, and an SSA is performed on the subsequent model output to compute sensitivity indices.

We choose to focus on $12$ physically distinct locations in the Arctic (Tab.~\ref{Sobol_Locations}) so as to characterise the response in various Arctic regions. These $12$ locations with $1,536$ simulations over the period of 2011--2022 represent approximately $200,000$ years of Icepack model data with which we perform our SSA of the Icepack model to its melt pond parameters. 

\begin{table}
\begin{center}
\resizebox{0.5\textwidth}{!}{%
\begin{tabular}{| c| c| c| c|} 
\hline
\multicolumn{4}{|c|}{\textbf{Locations used for Sobol sensitivity analysis}} \\
\hline
Location & Region & Latitude & Longitude \\ [0.5ex] 
\hline
 1 & Beaufort Sea & $72^\circ$ N & $137^\circ$ W \\
 2 & Chukchi Sea & $69^\circ$ N & $172^\circ$ W \\ 
 3 & East Siberian Sea & $72^\circ$ N & $163^\circ$ E \\ 
 4 & Laptev Sea & $76.3^\circ$ N & $125.6^\circ$ E \\  
 5 & Kara Sea & $77^\circ$ N & $77^\circ$ E \\  
 6 & Barents Sea & $75^\circ$ N & $40^\circ$ E \\ 
 7 & Greenland Sea & $76^\circ$ N & $8^\circ$ W \\ 
 8 & Fram Strait & $78^\circ$ N & $0^\circ$ E \\ 
 9 & Lincoln Sea & $82^\circ$ N & $58^\circ$ W \\ 
 10 & Baffin Bay & $74^\circ$ N & $68^\circ$ W \\ 
 11 & Hudson Bay & $60^\circ$ N & $86^\circ$ W \\ 
 12 & Arctic Ocean & $90^\circ$ N & $0^\circ$ E \\ 
 \hline
\end{tabular}
}
\end{center}
\captionof{table}{
\label{Sobol_Locations}
The locations used for our SSA of the Icepack model represent $12$ physically distinct areas of the Arctic.}
\end{table}

\begin{figure}
\centering
\includegraphics[width=0.9\textwidth]{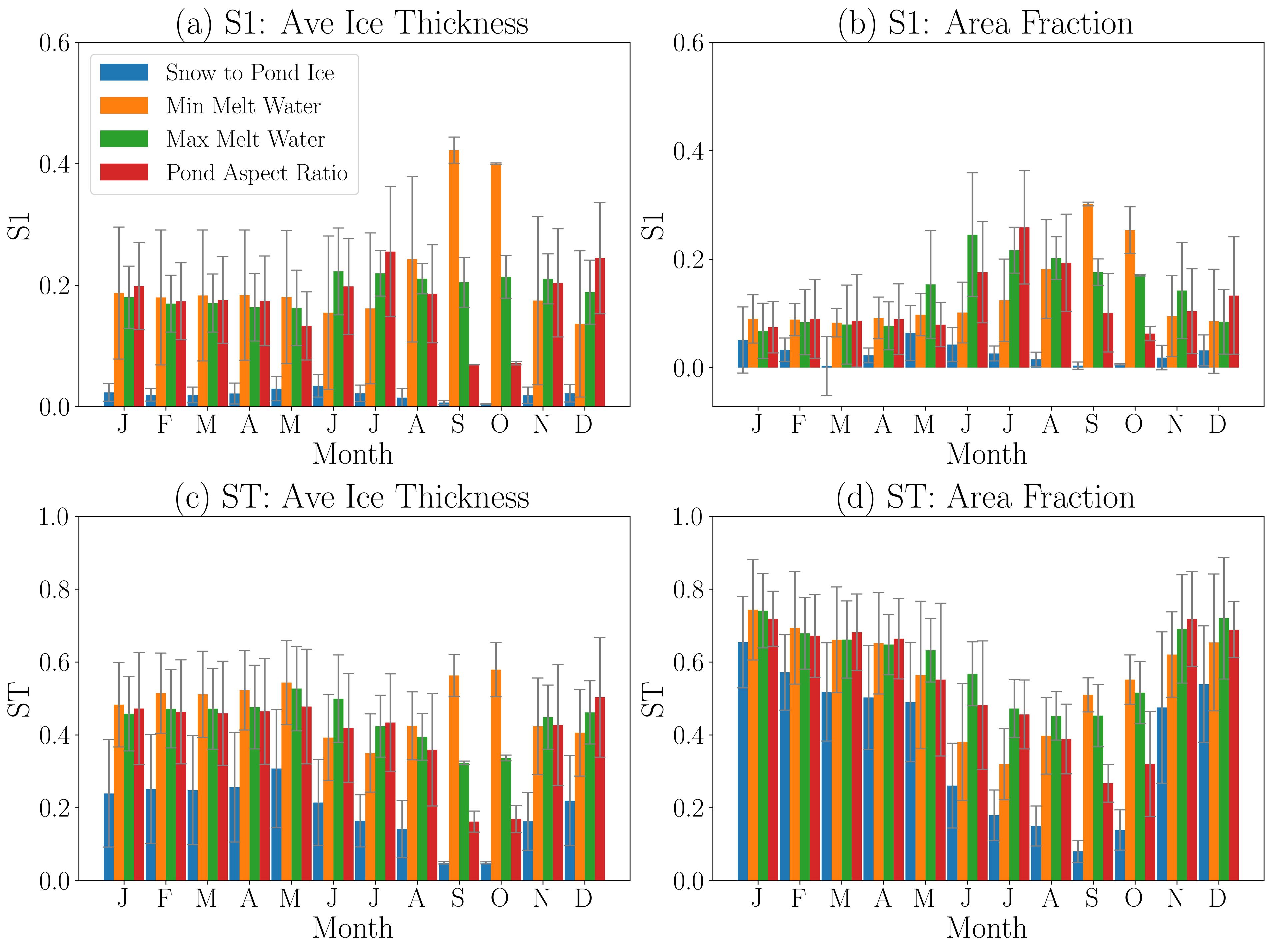}
\caption{Monthly first and total order Sobol indices relative to the 5 parameters in Tab.~\ref{mp_params}, averaged over the period 2012-2022, and averaged over all the $12$ locations in Tab.~\ref{Sobol_Locations}. Whisker plots on the monthly averages show the standard deviation of total order Sobol indices over $12$ locations.}
 \label{fig:S1_ST}
\end{figure}

Arguably the two most fundamental variables to consider when modelling sea ice are the sea ice thickness and area fraction, effectively representing the total volume of sea ice in a given region. Figure~\ref{fig:S1_ST} shows the results of SSA: the first order and total order Sobol indices for these key sea ice variables. Therefore in Fig.~\ref{fig:S1_ST} we are analysing the sensitivity of the Icepack model's variables related to sea ice volume with respect to its melt pond parameters.

Figure~\ref{fig:S1_ST} displays the monthly averaged first and total order sensitivity values averaged over the whole simulation period (2011-2022). We show the sensitivity to four over the five parameters in the melt pond parametrisation of Icepack (cf Tab.~\ref{mp_params}): the parameter ``snow depth of transition to bare ice'' ($hs0$ in Tab.~\ref{mp_params}) has no effect in the current scheme, and is thus excluded from the analysis.  
The Sobol indices are averaged across the $12$ locations to produce a single average first and total Sobol indices, and the vertical grey lines indicate the standard deviation over the $12$ geographical locations. Thus Fig.~\ref{fig:S1_ST} quantifies the sensitivity of the simulated ice thickness and are fraction to four parameters, expressing also its seasonality and the geographical variability. 

Figure~\ref{fig:S1_ST} shows that there exists a difference in the sensitivity of the model to each parameter - i.e. not all parameters are equal in their influence. The parameter \emph{hs1} (marking the snow depth of transition to pond ice, in blue) has the least influence across all Sobol indices and variables. The other parameters are associated with the minimum and maximum amount of available melt water to be added to melt ponds, and the last governs the pond aspect ratio (i.e. the area/height ratio of a pond of given volume). For a given volume of pond, altering the aspect ratio to make a smaller depth but bigger surface area pond would cause more absorption of incoming solar radiation, and therefore further melting. By contrast, a deep, narrow pond of the same volume would absorb incoming solar radiation, but would leave larger areas for exposed bare ice to reflect away the incoming solar radiation. Our SSA supports the fact that the amount of melt water added to a melt pond and the surface area of the pond are fundamental and obviously influential features when modelling melt ponds. 

The qualitative differences in first order and total order indices between variables (e.g. Fig.~\ref{fig:S1_ST}a-c, and Fig.~\ref{fig:S1_ST}b-d) indicate that the Icepack model's sensitivity to all of its melt pond parameters (as measured by ST) is more complicated than the effects of varying parameters individually (measured by S1). By construction the total order indices will always be larger than first order indices, as the total contains the sensitivity to that parameter alone, and higher order interactions. For instance, the parameter with the smallest S1 in Fig.~\ref{fig:S1_ST}a and b) (\emph{hs1}) has a much larger influence on the models output when considering interactions with other parameters (compare S1 and ST of this parameter in Fig.~\ref{fig:S1_ST}a and Fig.~\ref{fig:S1_ST}c). The same holding true for all parameters suggests that individually fine-tuning parameters in a sea ice model may not necessarily be the best approach, given this would neglect and ignore higher order effects. Whilst the first order effects are important, they neglect higher order interactions and the ``total contribution'' of each parameter towards the model output. Thus we will focus the remainder of this analysis predominantly on the total order Sobol indices.

The sensitivity of Icepack outputs to all of these parameters varies according to the time of year. 
We observe a decreasing ST of the Icepack model's area fraction to its melt pond parameters (Fig.~\ref{fig:S1_ST}d) as it enters the summer. This may be due to large (and increasing) areas in the Arctic having ice free periods as the summer approaches, thus limiting the influence of the melt pond parameters: there will be less ice to melt. Nonetheless, not all areas in the Arctic will fully melt to create an ice free area, and thus we can see that even throughout the summer the average values for any index and variable are non-zero. This rationale is further supported by the large standard deviation over the locations (grey vertical lines), suggesting that while locations at lower-latitude might be free of ice, higher-latitudes ones still present ice.

\begin{figure}
\includegraphics[width=1.0\textwidth]{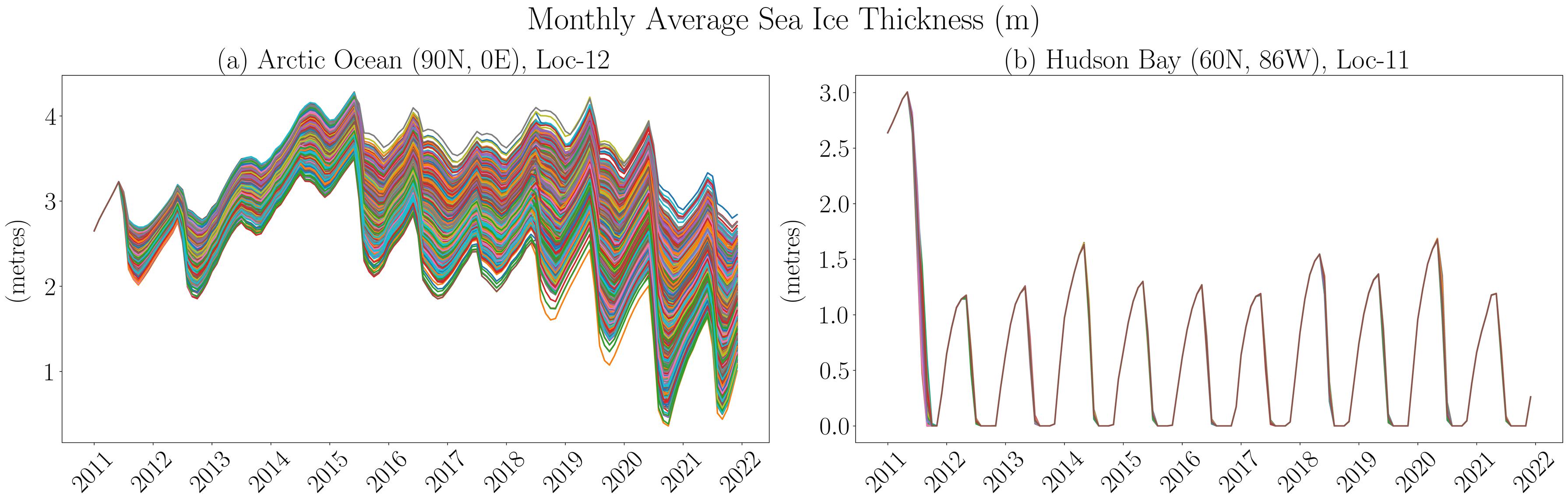}
\centering
\caption{Time series of monthly averaged sea ice thickness (metres) over the southernmost and northernmost locations chosen for SSA across all sampled parameter values. Each line represents an individual simulation in the 1,536 member ensemble at each location. }
\label{ait_hudson_arctic}
\end{figure}

To further unveil the geographical dependence of the model sensitivity to parameters, a manifestation of diverse and differently dominant physical process, we look at the model runs at two exemplar locations. The first is the Southernmost location: the Hudson Bay ($\#11$ in Tab.~\ref{Sobol_Locations}): it freeze over completely in winter but thaw in the summer. The second is the Northernmost location, the Arctic Ocean ($\#12$ in Tab.~\ref{Sobol_Locations}). Figure~\ref{ait_hudson_arctic} shows the time series of the monthly average sea ice thickness for each perturbed parameter member in the ensemble used for SSA. The difference between the two locations is huge. 
We see that not only does altering melt pond parameter values have substantially different effects depending on the location, but also that, for the Arctic Ocean location (panel a) the differences in predicted sea ice thickness can differ by $\approx 2-3$ metres. 

Overall our results reveal the Icepack model is highly sensitive to its melt pond parameters and that this sensitivity is both temporally and spatially varying.
This implies that parameters tested on various locations, and fine tuned to these locations, will generally not be optimal elsewhere. A parametrisation that could be tested, and fine tuned for stability, at one location, may have little influence on model output at this location. At the same time, these small fine tunings for stability may have a large influence on model output at another location. However, we can see here that small changes in parameter values that appear innocuous at one location could have enormous implications for sea ice thickness predictions at another location. That a small perturbation in these parameters can produce up to a few metres difference predicted sea ice thickness in the Arctic Ocean over approximately a decade is especially important for long term and climate change predictions. 

Climate model parameterisations, as the MPP in Icepack, are in general geographically and temporally invariant, ``one-size-fits-all'' approach. Ideally parametrisation should incorporate a spatio-temporal dependency and the complexity of processes ongoing in the Arctic that lead to the formation of melt ponds. What may seem relatively inconsequential uncertainty in one area, can lead to dramatically different sea ice predictions in another area.  
Can data-driven parametrisations emulate accurately physical parametrisation of the melt ponds in different geographical locations and seasons? Can they offer an alternative, computationally cheaper solution? Are they less sensitive to parameters' specifications?
Those are the questions addressed in the remaining of this paper. 

\section{Emulation of a melt pond parametrisation} \label{emulation_of_mpp}

We seek to understand if an emulator can learn and replace a physics-based MPP. We shall work under a ``perfect data'' assumption, in which our emulator is trained against exact (and complete) output of the MPP that it intends to replicate. While this assumption will be relaxed in follow-up studies where we will train our emulator against real data (see Sect.~\ref{disc_and_future} on our ongoing and future directions), this initial step is powerful as
it allows for a detailed understanding of the appropriate NN architecture, its capability to learn the relevant physical processes, and its computational stability. 

We will follow a two-stage experimental protocol. First, in the so called ``offline emulation'' (Sect.~\ref{ML-MPP_offline}) we aim at learning the physical MPP input-to-output mapping. Later, in the ``online emulation'' (Sect.~\ref{ML-MPP_online}) the ML-based MPP is plugged into the full sea-ice model Icepack, substituting the original physical MPP. This latter stage is substantially more challenging as it also looks at the computational stability of the ML-based parametrisation in the long run.   

\subsection{Dataset and neural network setup} 

\begin{figure}[!h]
    \centering
    \includegraphics[width=0.6\textwidth]{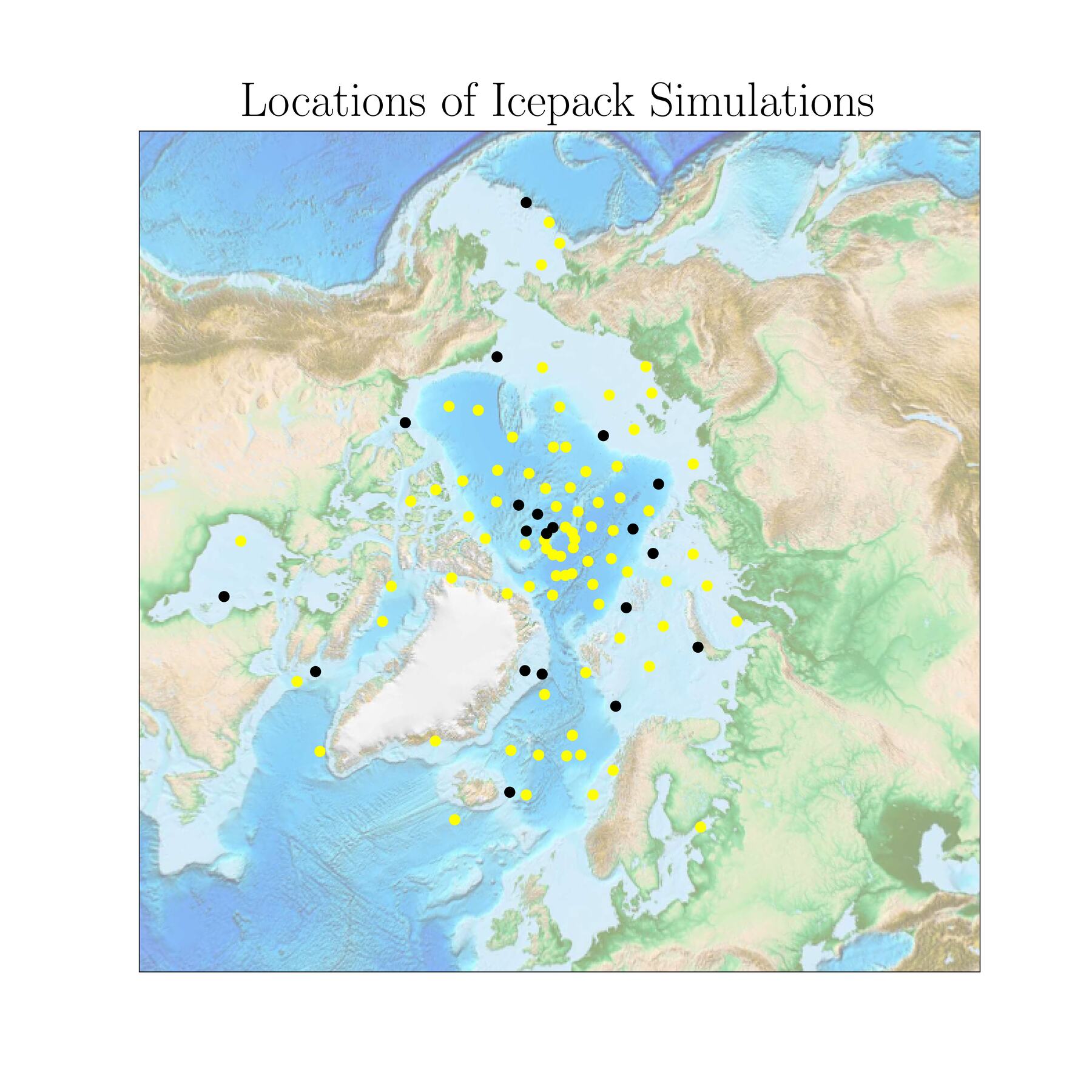}
    \caption{Locations of Icepack simulations, chosen at random from CFSv2 over its $1.875^{\circ} \times \approx 1.904^{\circ}$ resolution grid at latitudes between $60^{\circ}$N and $90^{\circ}$N. Yellow locations are those used as training data locations, a total of 80, whilst black locations represent those withheld for the testing stage ($20$ locations).}.
    \label{fig:icepack_locations}
\end{figure}

We create many simulations of the Icepack model, thus building a training dataset for NNs to learn the parametrisation from. Like with SSA, the forcing data we choose for Icepack comes from the the NCEP Climate Forecast System Version 2 (CFSv2). As opposed to our SSA study, there is no need to limit the database samples to an extraction from $12$ locations. Thus we choose $100$ randomly selected oceanic grid points over the Arctic circle and nearby latitudes in the CFSv2 dataset for our ML experiments (see Fig.~\ref{fig:icepack_locations}). The lowest occurrence of sea ice in the Northern Hemisphere occurs along the coast of Hokkaido, Japan \citep[$\approx 43^{\circ}$N,][]{Takahashi2011}. Accordingly, our 100 locations are randomly chosen from the CFSv2 dataset (i.e. over the $1.875^{\circ} \times \approx1.904^{\circ}$ resolution grid at latitudes between $60^{\circ}$N and $90^{\circ}$N). This yields mostly locations within the Arctic Circle whilst permitting training and testing locations in, for example, Hudson Bay, Canada, known for substantial sea ice coverage.

Input to the MPP comprises many model variables, such as surface temperature, level-ice area fraction, water fraction retained for melt ponds, rainfall rate, snow melt, top ice melt, depth difference for snow on sea ice and pond ice, and so on. The four output variables (e.g., the targets) calculated in the MPP are pond area, pond height, and frozen lid height (their calculated/outputted values at the previous timestep are also used as inputs to the subsequent timestep) and fraction of atmosphere/ice heat flux used to melt frozen lids which is calculated purely from the input variables anew at each timestep, therefore three of the four target variables are calculated explicitly using also their values at the previous timestep. The full list of inputs and targets is given in Tab.~\ref{tab:icepack_output_mpp}

\begin{center}
\begin{table}
\begin{center}
\begin{tabular}{| c | c |} 
\hline

\multicolumn{2}{|c|}{\bf Inputs/Features and Outputs/Targets in the MPP of Icepack} \\
\hline
\textbf{Features} & \textbf{Description} \\ [1ex]
\hline 
rfrac & Water fraction retained for melt ponds \\ [1ex] 
\hdashline
meltt & Top melt rate (m/s)\\ [1ex] 
\hdashline
melts & Snow melt rate (m/s) \\ [1ex] 
\hdashline
frain  & Rainfall rate (kg/m$^2$/s) \\ [1ex] 
\hdashline
Tair  & Air temperature (K) \\ [1ex] 
\hdashline
fsurfn & Atmosphere-ice surface heat flux  ($W/m^2$)  \\ [1ex] 
\hdashline
dhs & Depth difference for snow on sea ice and pond ice (m) \\ [1ex] 
\hdashline
aicen & Ice area fraction \\ [1ex] 
\hdashline
vicen & Ice volume (m) \\ [1ex] 
\hdashline
vsnon & Snow volume (m) \\ [1ex] 
\hdashline
Tsfcn & Surface temperature ($^\circ$ Celsius) \\ [1ex] 
\hdashline
alvl & Level-ice area fraction \\ [1ex] 
\hdashline
 $\textup{apnd}^{\dag}$ & Melt pond area fraction \\
\hdashline
 $\textup{hpnd}^{\dag}$ & Melt pond height/depth \\ 
\hdashline
 $\textup{ipnd}^{\dag}$ & Melt pond refrozen lid thickness \\
 \hline
 \textbf{Targets} & \textbf{Description} \\ [1ex]
\hline
 apnd & Melt pond area fraction \\
\hdashline
 hpnd & Melt pond height/depth \\ 
\hdashline
 ipnd & Melt pond refrozen lid thickness \\
\hdashline
ffrac & Fraction of fsurfn to melt ipnd \\ 
\hline
\end{tabular}
\end{center}
\captionof{table}{All variable features (inputs) and targets (outputs) of the melt pond parametrisation in the level-ice scheme with the ``mushy thermodynamics'' option. Constants are also supplied to the MPP, such as \emph{nilyr} (the number of ice layers), \emph{dt} (time step in seconds),  \emph{ktherm} (thermodynamics scheme option), and not all variables supplied to the MPP may be used for all schemes. For example, as the mushy thermodynamics scheme handles flushing, \emph{dpscale}, the alter e-folding time scale for flushing, and \emph{qicen} (the ice layer enthalpy) which would used for explicit calculations for the brine permeability of the ice, are not required. Our table comprises all input that is relevant for the level-ice melt pond parametrisation using the mushy thermodynamics scheme, using all default parameter options in the Icepack model - see \citet{Hunke2021} for more details. For each timestep Icepack makes 5 calls to the MPP, one for each ice category. \\
$\textup{ }^{\dag}$ implies the value from the previous timestep.}
\label{tab:icepack_output_mpp}
\end{table} 
\end{center}

We train four NNs, one per target, rather than a single multi-target NN. The processes that lead to the different MPP targets appear to be quite different, representing an inherent challenge for a single NN to reproduce them all simultaneously. Thus the benefit of having one NN to train for all targets is foregone in favour of greater flexibility.
As common practice the full dataset is split into a training, validation and test data so as to understand the performance of the model on data that was not used in the training: some data is held back and we ask the model to make predictions for that period. For time series data like we have in this study, data are usually split time-wise \citep[e.g., walk-forward][]{Assaad2021}. 
As we deal with many time series over multiple locations, we have instead split the data by location, i.e. NNs are trained on full time series data from Icepack simulations in the period 2012--2021 (inclusive) and over $80$ locations across the Arctic (yellow dots in Fig.~\ref{fig:icepack_locations}). The $20$ test locations (black dots in Fig. ~\ref{fig:icepack_locations}) are used to assess if our emulator has learned the underlying physics and can emulate sea ice melting at other regions in the Arctic. Results using this spatial partition of training and testing data yields close agreement (not shown) to a more standard time partition in which the data 2011--2019 were used for training and 2020--2021 used as test data.

\begin{figure}[H]
\includegraphics[scale=0.11]{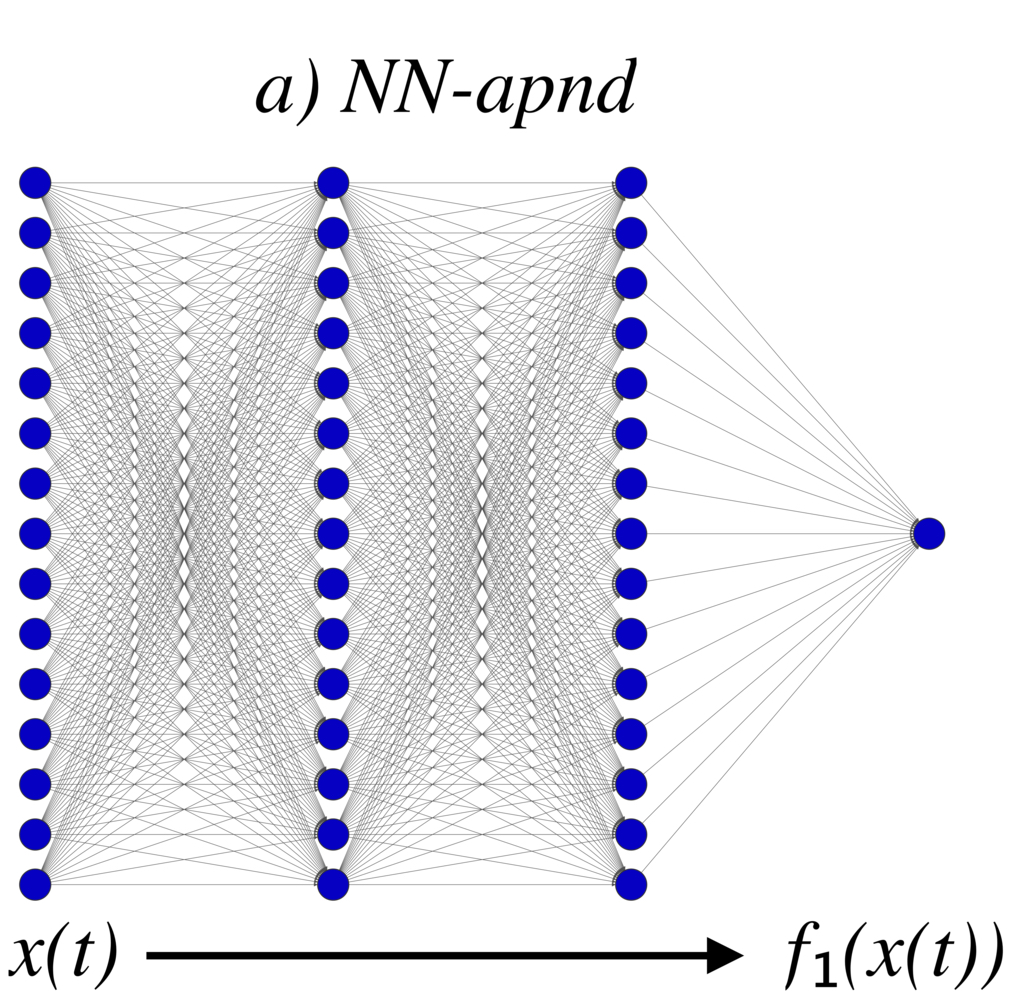}
\includegraphics[scale=0.11]{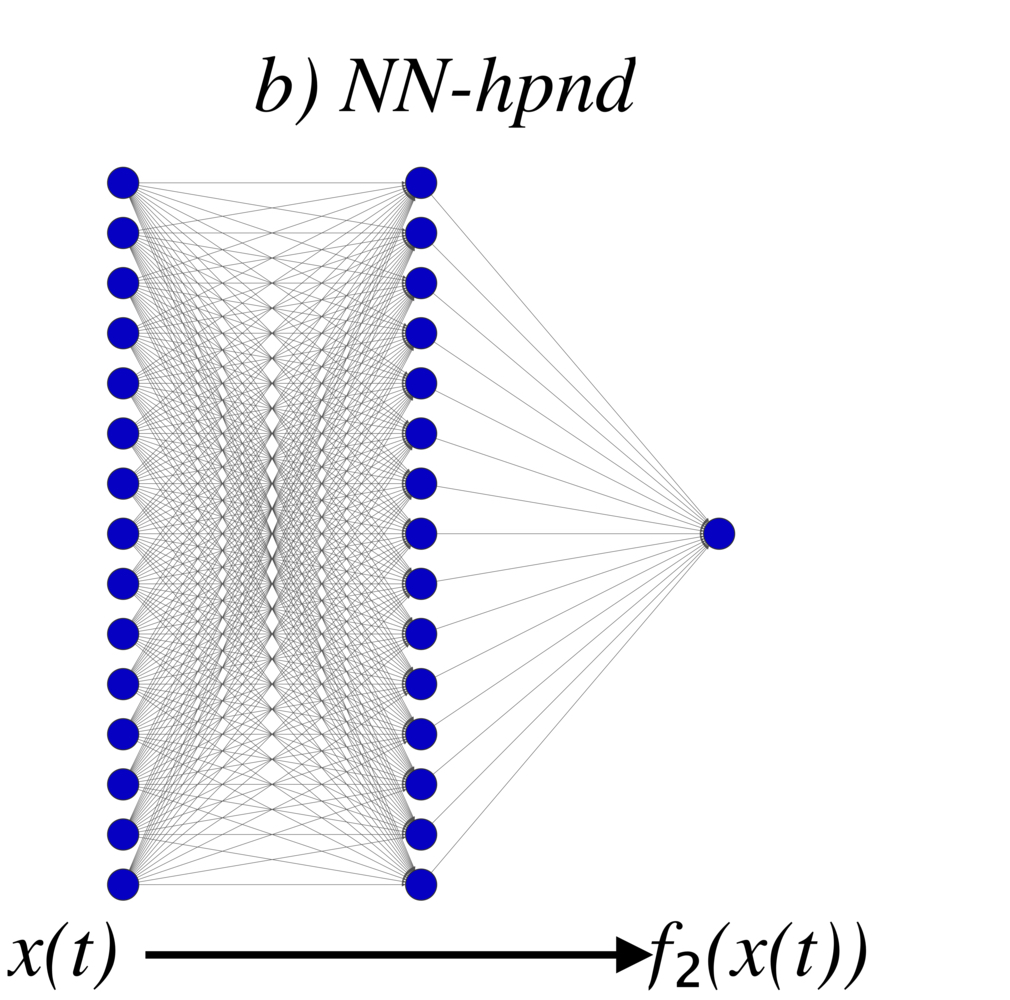}
\includegraphics[scale=0.11]{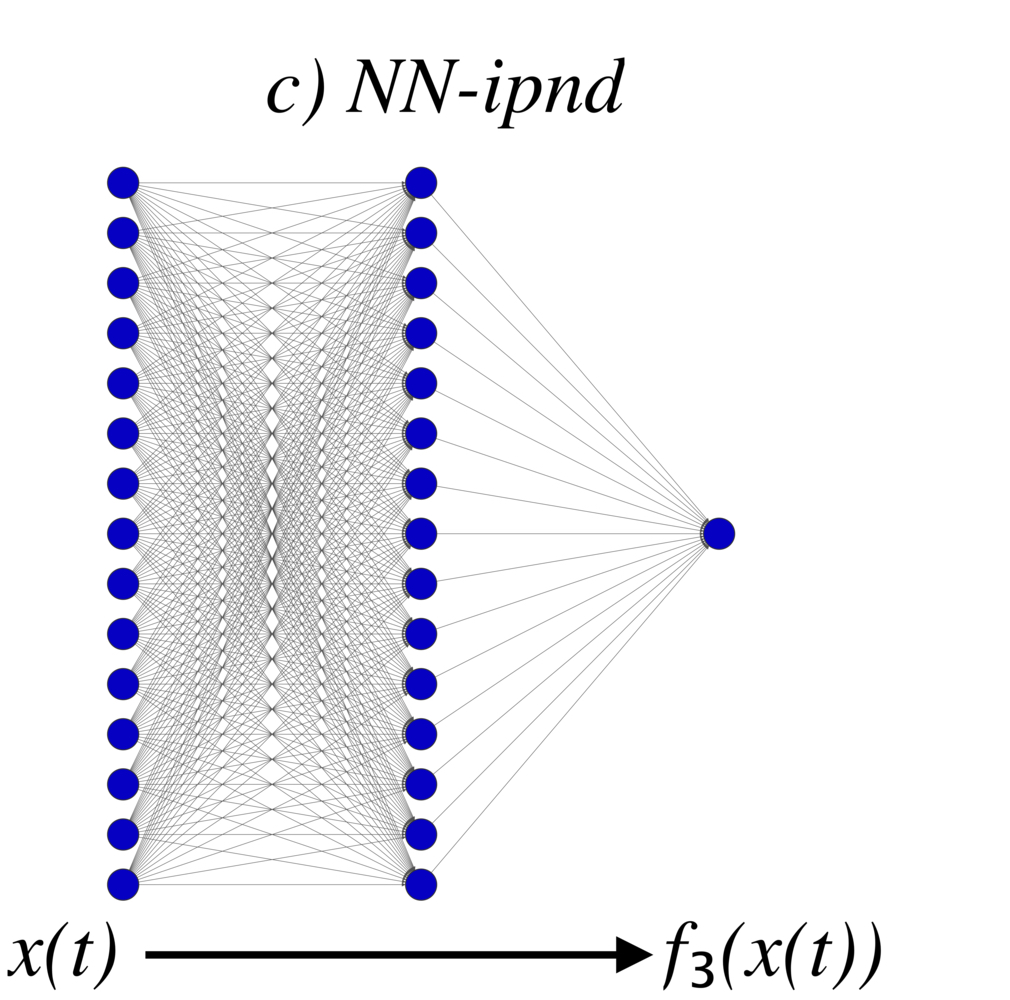}
\centering
\caption{Representation of the architecture of the neural networks for pond area, pond height, frozen lid (``ipond'') height and fraction of atmosphere-ice heat flux used to met frozen lids. All networks use the sigmoid activation function as the output layer, which also comprises the activation functions in the hidden layers for all but the final neural network, which uses the swish activation function for the hidden layers. 50 epochs are used to train the NNs.}
\label{fig:nn_architecture}
\end{figure}

Given the importance of hyperparameters in NNs performance, we initially performed a hyperparameter optimisation \citep{Hutter2014} using  Hyperband \citep{Li2018}, a novel bandit-based approach: a form of random search that is sped up through early stopping and adaptive resource allocation. We  searched over a hyperparameter space consisting of the number of nodes, layers, and types of activation functions. Hyperband selected four deep NNs that returned slightly better scores for offline training. Performance was assessed relative to a validation set where a training, validation and test set were split by $60$, $20$ and $20$ locations. These deep NNs caused Icepack to become slightly unstable, thus we reduced the number of hidden layers and trained shallower NNs on a merged training and validation set (all $80$ yellow locations in Fig.~\ref{fig:icepack_locations}). We choose these shallow NNs subjectively, as a trade-off between accuracy and complexity. 
Hereafter, all results are relative to these latter shallow NNs that are illustrated in Fig.~\ref{fig:nn_architecture}. These NNs are trained on the $80$ locations over $50$ epochs, and we report their performance finally relative to the test locations, which are held back from the model selection process. 

\subsection{ML-based melt ponds parametrisation - Offline emulation} \label{ML-MPP_offline}

\begin{figure}[!h]
    \centering
    \includegraphics[width=1.0\textwidth]{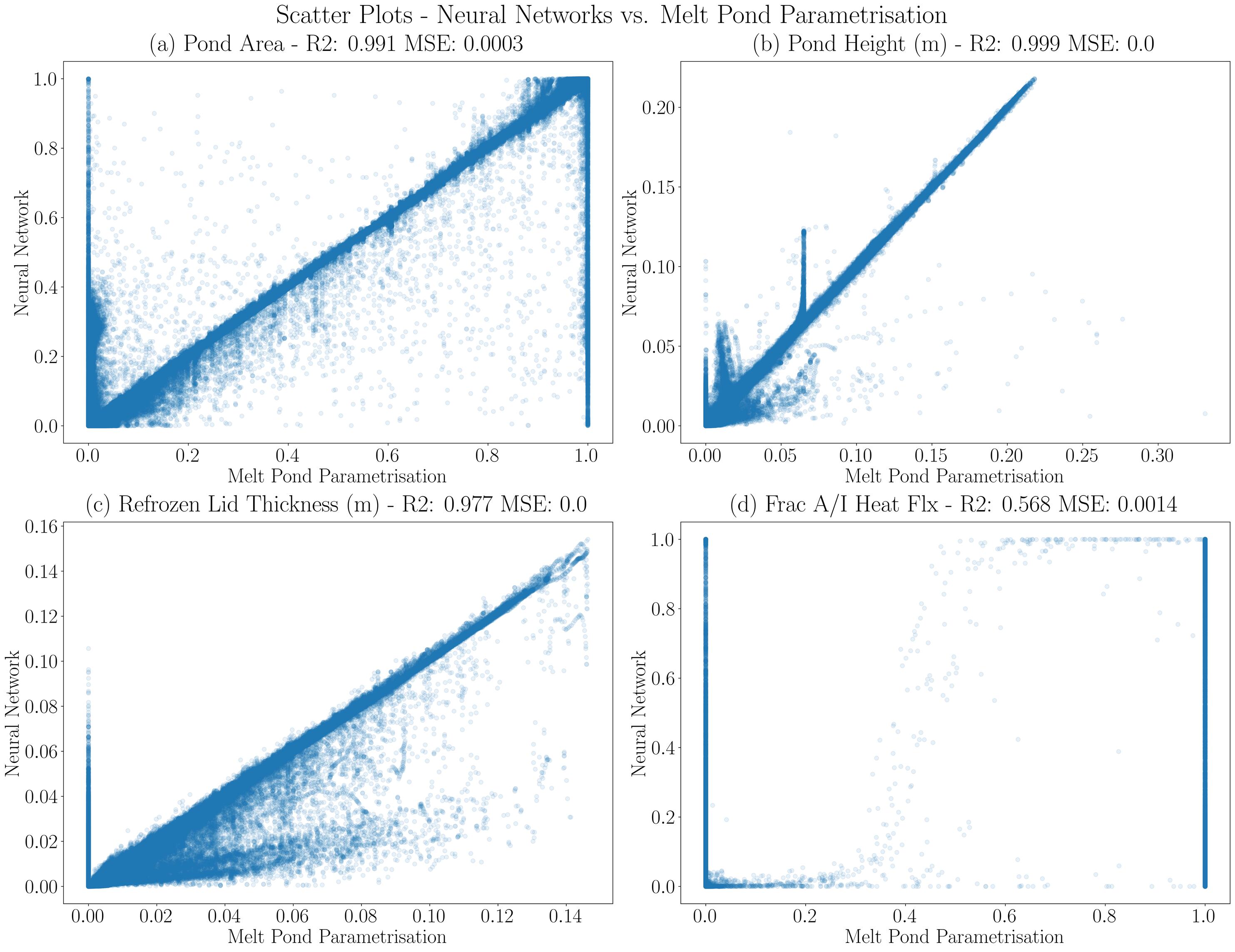}\\
    \caption{Scatter plots of the four target variables comparing their true values and the predicted values as given by the neural networks.}
    \label{fig:offline_scatter}
\end{figure}

Figure~\ref{fig:offline_scatter} shows the scatter plots of the predicted values against the targets (the values predicted by the physical MPP), together with the corresponding R2 score and mean square error (MSE). The figure shows that overall the NNs have been able to learn the feature-to-target relationship in the physical MPP. The performance is very good for three targets, pond area, pond height, and the ``ipond height'', or height of the ice lid on a frozen pond, achieving extremely high R2 scores ($> \approx 0.97$ for all three targets). For the fourth target, ``Frac A/I Heat Flx'', the performance is of medium skill (R2$=0.568$), but this is indeed a much more complex variable representing the fraction of atmosphere/ice heat flux calculated to melt the frozen lids. 

The \emph{ffrac} target is almost always zero or one, where samples with $\emph{ffrac}=1$ making up $\approx 0.25\%$ of the dataset, samples where $0<\emph{ffrac}<1$ $\approx 0.04\%$ of the dataset, and all other samples being where $\emph{ffrac}=0$. Thus it represents an extreme imbalanced learning problem, and as \emph{ffrac} is almost always either zero or one, it is very close to an imbalanced binary classification problem. 
Traditional ML algorithms assume that the number of samples in each class is approximately equal \citep[e.g.,][]{Krawczyk2016}, and algorithms will be biased towards the majority group. As learning of \emph{ffrac} represents learning a nonlinear and extremely imbalanced problem, it differs substantially to the other three targets. Nonetheless, we shall see in Sect.~4.3 that the skill of the NNs is sufficient for online emulation of our parametrisation. 

From the point of the view the NNs' goal of learning the input-output relation in the physical MPP, the difficulty in the variable \emph{ffrac} relies also in its strong temporal variability. By contrast, the other three variables directly depend on their values at the previous timestep (an input to the MPP, see Tab.~\ref{tab:icepack_output_mpp}), making it possible to predict their future value with sufficient accuracy, even with a mere linear regression; see Tab.~\ref{tab:regression_scores}.
Notwithstanding, table~\ref{tab:regression_scores} shows also that, whilst linear regressions do an excellent job in capturing two of the four target variables (pond area and height), NNs perform better than linear regressions for all target variables. In particular, for the thermodynamic-flavoured target variables ({\it ffrac}), NNs show substantial improvements where linear regressions show almost no predictive ability. 

\begin{center}
\centering
\begin{tabular}{| p{1.8cm}| p{1.8cm}| p{1.8cm}|  p{1.8cm}| p{1.8cm}|} 

\hline
\multicolumn{5}{|c|}{\textbf{Linear Regression and Neural Network Test Data Scores}} \\
\hline
\multicolumn{5}{|c|}{\textbf{Mean Square Error}}
\\
\hline
 & \centering apnd & \centering hpnd & \centering ipnd & ffrac \\
\hline
Lin Reg & 6.03$\times10^{-4}$ & 1.45$\times10^{-6}$& 5.56$\times10^{-6}$ & 3.07$\times10^{-3}$ \\ [0.5ex] 
\hdashline
NN & \textbf{3.23$\times\textbf{10}^{\textbf{-4}}$} & \textbf{6.24$\times\textbf{10}^{\textbf{-7}}$} & \textbf{1.65$\times\textbf{10}^{\textbf{-6}}$} & \textbf{1.35$\times\textbf{10}^{\textbf{-3}}$}\\
\hline
\multicolumn{5}{|c|}{\textbf{R2 Score}}\\
\hline
Lin Reg & 0.983 & 0.997 & 0.923 & 0.017 \\ [0.5ex] 
\hdashline
NN & \textbf{0.991} & \textbf{0.999} & \textbf{0.977} & \textbf{0.568} \\
\hline
\end{tabular}
\captionof{table}{Scores of linear regressions and neural networks for each of the four target variables on the test data. Errors are calculated relative to the target variables as given by the MPP itself. \emph{apnd} and \emph{ffrac} are unitless variables, \emph{hpnd} and \emph{ipnd} are measured in units of metres. \emph{(bold numbers indicate better performance)}}
\label{tab:regression_scores}
\centering
\end{center}

\subsection{ML-based melt ponds parametrisation -- Online emulation} \label{ML-MPP_online}

The ``online emulation'' challenge consists in substituting the original physics-based MPP in the full sea-ice model Icepack with the ML-based MPP and in studying the new ML-based Icepack accuracy and stability. For completeness and for the sake of understanding the nature and complexity of the processes leading to melt ponds, we compare the original Icepack, hereafter referred to as {\it Icepack-Phys}, with our nonlinear NN-based MPP, {\it Icepack-NN}, and with {\it Icepack-LR} whereby the MPP is based on linear regression.  
As for the offline emulation, experiments run for the period of 2011-2021, over the 20 test locations displayed in Fig.~\ref{fig:icepack_locations}; performance is assessed with regards to locations that both LR- and NN-based emulators have not seen during the training. 

The linear regression optimised for the offline learning, and presented in Tab.~\ref{tab:regression_scores}, could not be directly used in the online setting because, being unbounded, it often predicts statistically sound but nonphysical values, such as area fractions of melt ponds greater than one, or less than zero.
While this already points to the inherent limitation of standard LR without constraint, for the sake of the comparison, in the experiments that follow we cap the LR output, to only yield physically plausible values.
This entails coercing all four target variables with a minimum predicted value of zero, and both \emph{apnd} and \emph{ffrac} with a maximum value of one. For instance, if LRs predict a negative pond height, we set the pond height to zero.  
It is worth noting that NNs are bounded by construction thanks to the use of the sigmoid activation function for all outputs. Very similar results (not shown here) occur with alternative activation functions, such as ReLU and tanh. 

\begin{figure}
\centering
\includegraphics[width=1.0\textwidth]{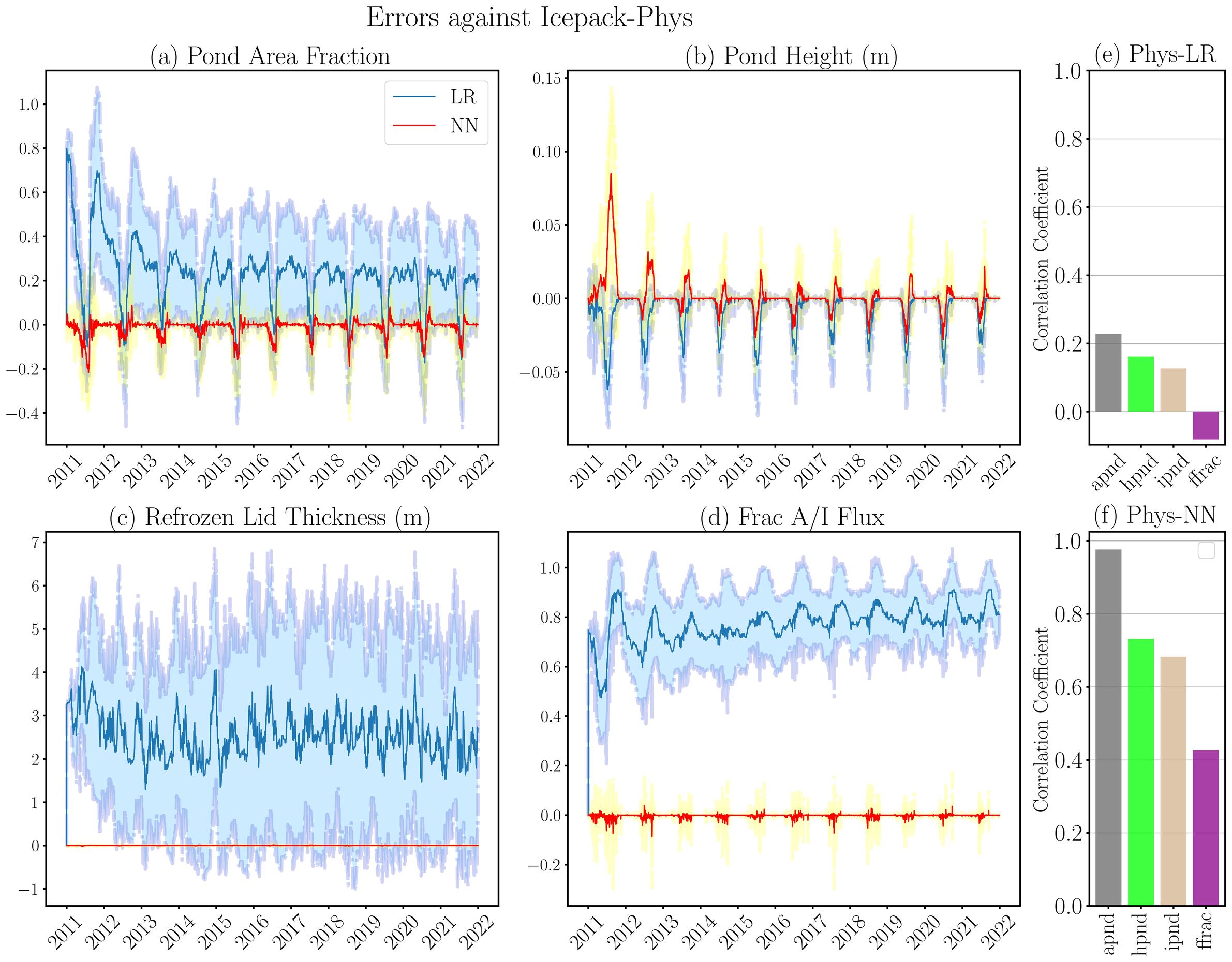}
\caption{Error time-series in the four MPP output variables: (a) Pond area fraction (apnd), (b) Melt pond height (hpnd), (c) Melt pond refrozen lid thickness (ipnd) and (d) Fraction of atmosphere-ice net heat flux at the surface used to melt (ffrac). Values are computed against Icepack-phys and refer to Icepack-LR (blue) and Icepack-NN (red). Solid lines represent the daily averaged values across all simulations (over all geographic locations), whilst blue (Icepack-LR) and yellow (Icepack-NN) shading above and below represents one standard deviation either side of the mean, respectively. Pearson correlations of the daily mean values of the four target variables between Icepack-phys and Icepack-LR are given in panel e and for Icepack-NN in panel f.}
\label{fig:error_timeseries_mpp}
\end{figure}

Figure~\ref{fig:error_timeseries_mpp} shows the error time-series of Icepack-LR and Icepack-NN; values are computed relative to Icepack-Phys. In contrast to the very accurate prediction of Pond area fraction (panel a) of Icepack-NN, Icepack-LR drifts rapidly away from Iceapack-Phys, simulating unrealistic very large pond area fractions, and near year round substantial pond coverage. 
Pond heights (panel b) in Icepack-LR are substantially different from Icepack-Phys, and the results suggest overall that Icepack-LR represents a state of very large areas of ponds that are covered with a frozen lid. 
Correlations between Icepack-Phys and Icepack-LR (panel e)) are not significantly correlated. This highlights the poor representation of melt ponds that can be achieved using linear regressions which in turn manifests further the nonlinear characters of the processes.

\begin{figure}
\centering
\includegraphics[width=1.0\textwidth]{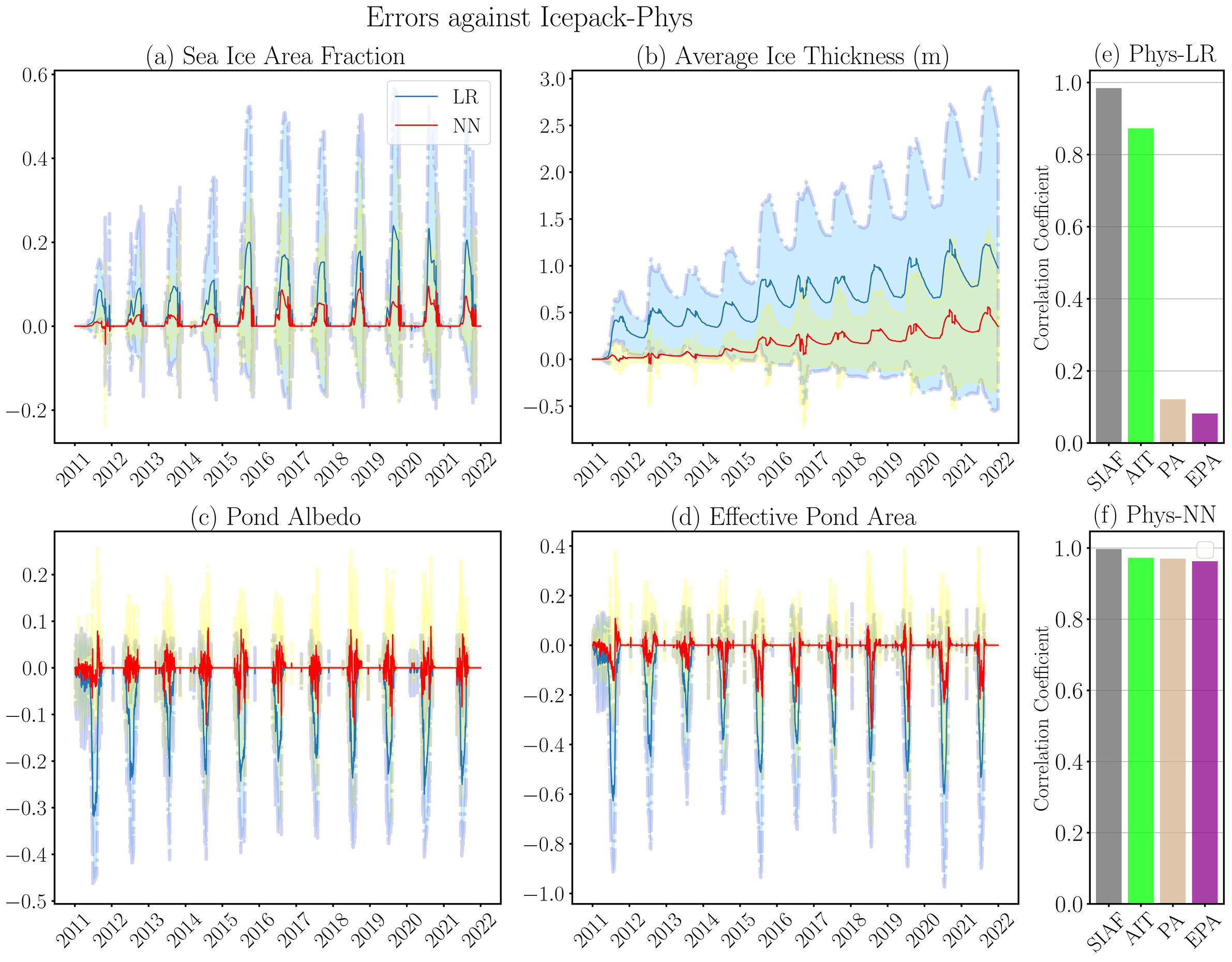}
\caption{Time series of error for linear regressions and neural networks, both compared to the physical MPP for key output variables. Shading represents one standard deviation width above and below the mean solid lines. (NB: SIAF $=$ ``Sea ice area fraction''; AIT $=$ ``Average Ice Thickness''; PA $=$ ``Pond Albedo''; EPA $=$ ``Effective Pond Albedo'')}
\label{fig:error_timeseries_itd}
\end{figure}

Figure~\ref{fig:error_timeseries_itd} has the same format and content as Fig.~\ref{fig:error_timeseries_mpp}, except that rather than focusing on MPP outputs, it displays error timeseries of key sea-ice variables simulated in Icepack: sea-ice area fraction ({\it SIAF}; panel a), average ice thickness ({\it AIT}; panel b), pond albedo ({\it PA};
panel c) and, effective pond area ({\it EPA}; panel d). Average correlations are given in panels e and f.  

Figure~\ref{fig:error_timeseries_itd} reveals important differences between Icepack-NN and Icepack-LR. Icepack-NN's errors in sea ice area fraction (Fig.~\ref{fig:error_timeseries_itd}a) are fairly small, with Icepack-NN on average slightly overestimating the sea ice area fraction. This is due to a slightly smaller area of melt ponds in summer meaning there is less melting occurring and less warming/absorption of incoming downwelling radiation to further warm the sea ice and reduce sea ice fraction. Icepack-LR errors, in these summer months are two to three times larger. Icepack-LR does a substantially worse job at representing melt ponds than Icepack-NN displaying very little summertime melt/feedback. 

The implications of this behaviour can be seen whenever the average ice thickness are too large relative to Icepack-Phys and it begins already in the first summer of Icepack-LR simulations. By contrast, Icepack-NN has small errors in the first year (Fig.~\ref{fig:error_timeseries_itd}b), in either average ice thickness or sea ice area fraction. This means that Icepack-NN is able to generate melt ponds as it does not suffer from a comparative lack of summer sea ice melting and does not have an immediate build up of ice thickness relative to Icepack-Phys (cf the $~0.5$~metres mean error for Icepack-LR within one year of emulation).

Figures~\ref{fig:error_timeseries_itd}c and \ref{fig:error_timeseries_itd}d show important variables' evolution for the radiative scheme in Icepack: pond albedo and effective pond area. On the one hand, errors for both are relatively small for Icepack-NN indicating that the emulator is able to produce melt ponds when run in Icepack fairly consistently with the physical MPP. On the other hand, Icepack-LR does not perform equally well. For instance, Fig.~\ref{fig:error_timeseries_itd}d shows large areas of melt ponds that are covered with frozen lids (i.e. almost fully covered), whilst Icepack-Phys has positive values for effective pond area in summer annually. Furthermore, we see annually negative error values for ``effective pond area'' (which corresponds to the exposed pond area, i.e. the area of a melt pond not covered with a frozen lid, but exposing open water) of Icepack-LR relative to Icepack-Phys, indicative of a very small or no effective pond area. Effective pond area in Icepack-LR in winter is the same as Icepack-Phys: to incoming radiation above it, ice is the same as a pond covered with ice. Accordingly, this yields a substantially increased average sea ice thickness of approximately $1$~metre on average by the end of the simulations. 

Correlation plots in Fig.~\ref{fig:error_timeseries_itd} show that whilst there exist high correlations between Icepack-LR and Icepack-Phys for sea ice area fraction and average ice thickness (suggesting a large role of model forcing on retaining some relation between the Icepack model output) correlations are poor for melt pond related variables. In general linear regressions do a very poor job of being able to emulate a melt pond parametrisation, necessitating more powerful techniques than using Icepack-LR.

By contrast, Fig.~\ref{fig:error_timeseries_mpp} shows no such rapid increase and deviation of values in Icepack-NN relative to Icepack-Phys. Small to moderate errors appear in the summer/melt season, however essentially return back to zero ponds in the winter, unlike Icepack-LR. Icepack-NN slightly underestimates pond area fraction on average, with pond heights overestimated by only around $1$ to $5$~cm (Fig.~\ref{fig:error_timeseries_mpp}). Correlation coefficients and the error time series show in general there is a good agreement between Icepack-NN and Icepack-Phys in both Fig.~\ref{fig:error_timeseries_mpp} and Fig.~\ref{fig:error_timeseries_itd}. A slightly smaller pond area (and effective pond area) contributes to slightly larger average ice thicknesses. Some drift after approximately 4-5 years does occur in average ice thickness in Icepack-NN as a gradual slow increase in relative sea ice thickness. We understand this is caused by a slightly smaller representation of melt ponds in Icepack-NN as compared to Icepack-Phys. Importantly, Icepack-NN does not exhibit the rapid error shifts seen in Icepack-LR, and it attains error levels as small as fractions of those in Icepack-LR. The results suggest that NNs used here have the capability to emulate to a reasonable degree the MPP, with a slight underestimation of melt ponds, and indicate that NNs can be used to emulate a physical parametrisation where, for the most part, they do not cause any substantial drift or instabilities. 


In the next Section we perform feature selection on the NN-based MPP in Icepack-NN, to determine if a model with a reduced number of the most important inputs can still give a reasonable performance, laying foundations for an (even further) computationally cheaper, possibly observationally trained, emulator where relations between inputs and outputs are clear and physically interpretable.

\subsection{Feature selection using mutual information} \label{fs_using_mi}




Given widespread use, and powerful results in ML, such as the \citep[UAT, ][]{Hornik1989} which states that there exists a multilayer feedforward NN able to approximate any Borel measurable function to any desired degree of accuracy, use of ML techniques is widespread. This has lead to NNs that can predict a model or function, but it do not provide guides on how to construct it from, for example, a physical perspective or on its interpretability. Machine learning can therefore be ``quasi''-blindly used in many diverse problems across the sciences, leading to emulators that perform well but act as a ``black-box'' \citep[e.g.,][]{Camburu2020}, from where it is difficult to discern the underlying physical, societal, economical, and so on, drivers. 

Feature selection helps devising a minimal model by identifying the inputs that have the largest impact on the output, thus evidencing the physical mechanism of greater importance. A reduced model is easier to interpret and it is usually less computationally demanding. In this study, we perform feature selection using the {\it mutual information} approach on the training data for each MPP target relative to the input features. Mutual information \citep{Shannon1948} measures the amount of information one (or group of) random variable(s) contains about another. In essence, it measures the dependency existing between two (groups of) continuous or discrete random variables.  

Let $X$ and $Y$ be random variables whose probability density functions are $P_{X}$ and $P_{Y}$ and whose domains are $\mathscr{X}$ and $\mathscr{Y}$, respectively. Let us also define their joint probability density function $P_{({X,Y})}$. The mutual information between $X$ and $Y$ is defined by
\begin{equation}
    I(X;Y) = - \int_{\mathscr{X}} \int_{\mathscr{Y}} P_{(X,Y)} (x, y) \ln \left( \frac{P_{(X,Y)}(x, y)}{P_{X} (x) P_{Y} (y)} \right)  {\mathrm d}x{\mathrm d}y,
\end{equation}
\citep[e.g.,][]{Frenay2013}. Mutual information can thus be understood as the reduction of uncertainty on the values of $Y$ once $X$ is known, and vice-versa. It also has the key advantage of detecting non-linear relationships between variables, thus eliminating the need to make any assumptions about the linearity between features and targets.

\begin{figure}[!h]
    \centering
    \includegraphics[width=1.0\textwidth]{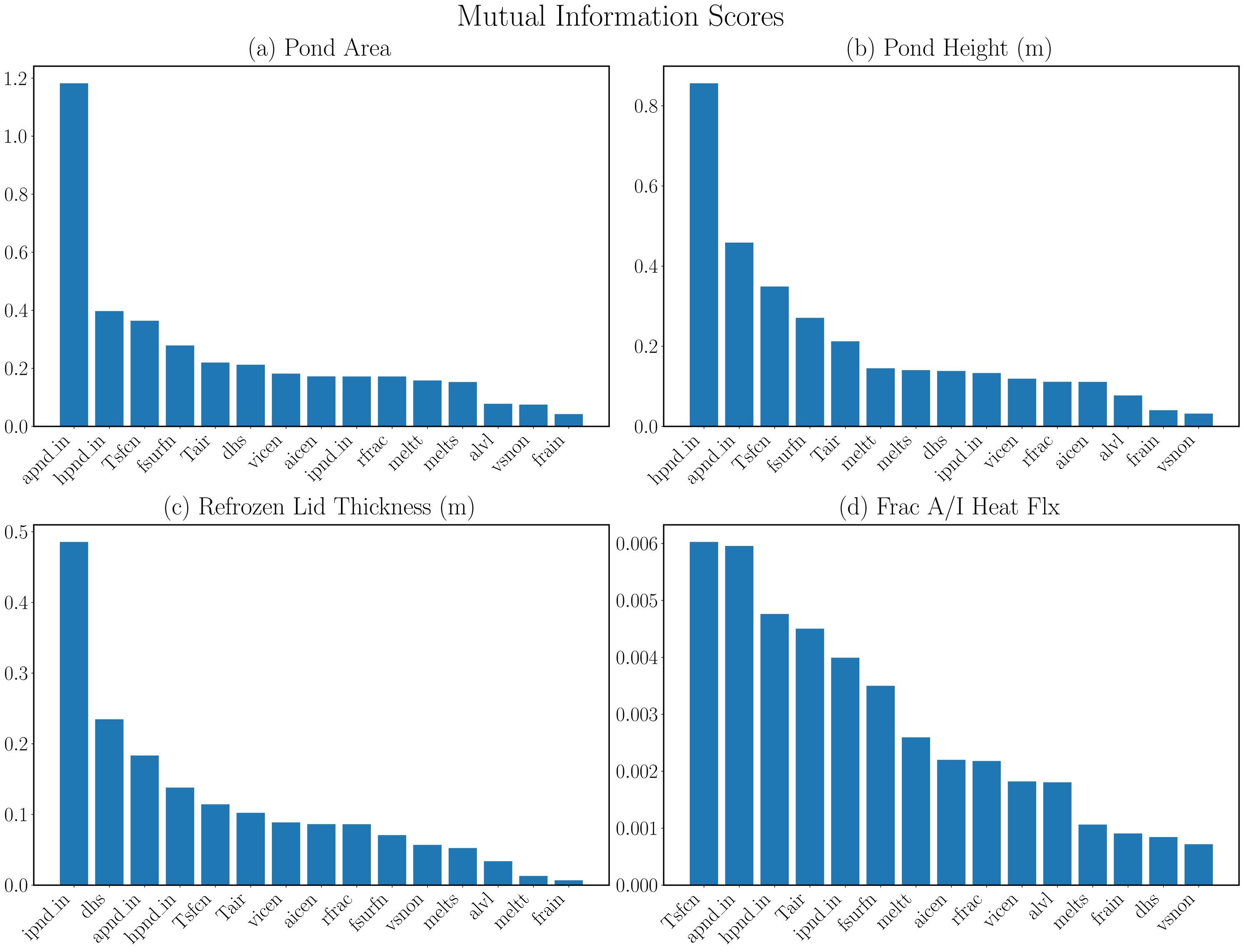}
    \caption{Feature selection with Mutual Information - features are ordered with respect to their mutual information scores for each target (apnd, hpnd, ipnd and ffrac). For the definition of each input variable, see Tab.~\ref{tab:icepack_output_mpp}.}.
    \label{fig:mi_feature_selection}
\end{figure}

Figure~\ref{fig:mi_feature_selection} shows the mutual information scores for each target variable with respect to the features. For pond area (Fig.~\ref{fig:mi_feature_selection}a), pond height (Fig.~\ref{fig:mi_feature_selection}b) and refrozen lid height, Fig.~\ref{fig:mi_feature_selection}c) the largest mutual information score is between that target and the value of that same variable at the previous timestep (recall that they are among the features; see Tab.~\ref{tab:icepack_output_mpp})): these play the most dominant roles in determining the value at next time step. By contrast, for the fraction of atmosphere/ice heat flux used to melt frozen lids (``Frac A/I H flux'', Fig.~\ref{fig:mi_feature_selection}d) a more gradual decline is seen suggesting that predictive power for this variable comes from the cumulative total of including more features.

\begin{figure}
\centering
\includegraphics[width=.6\textwidth]{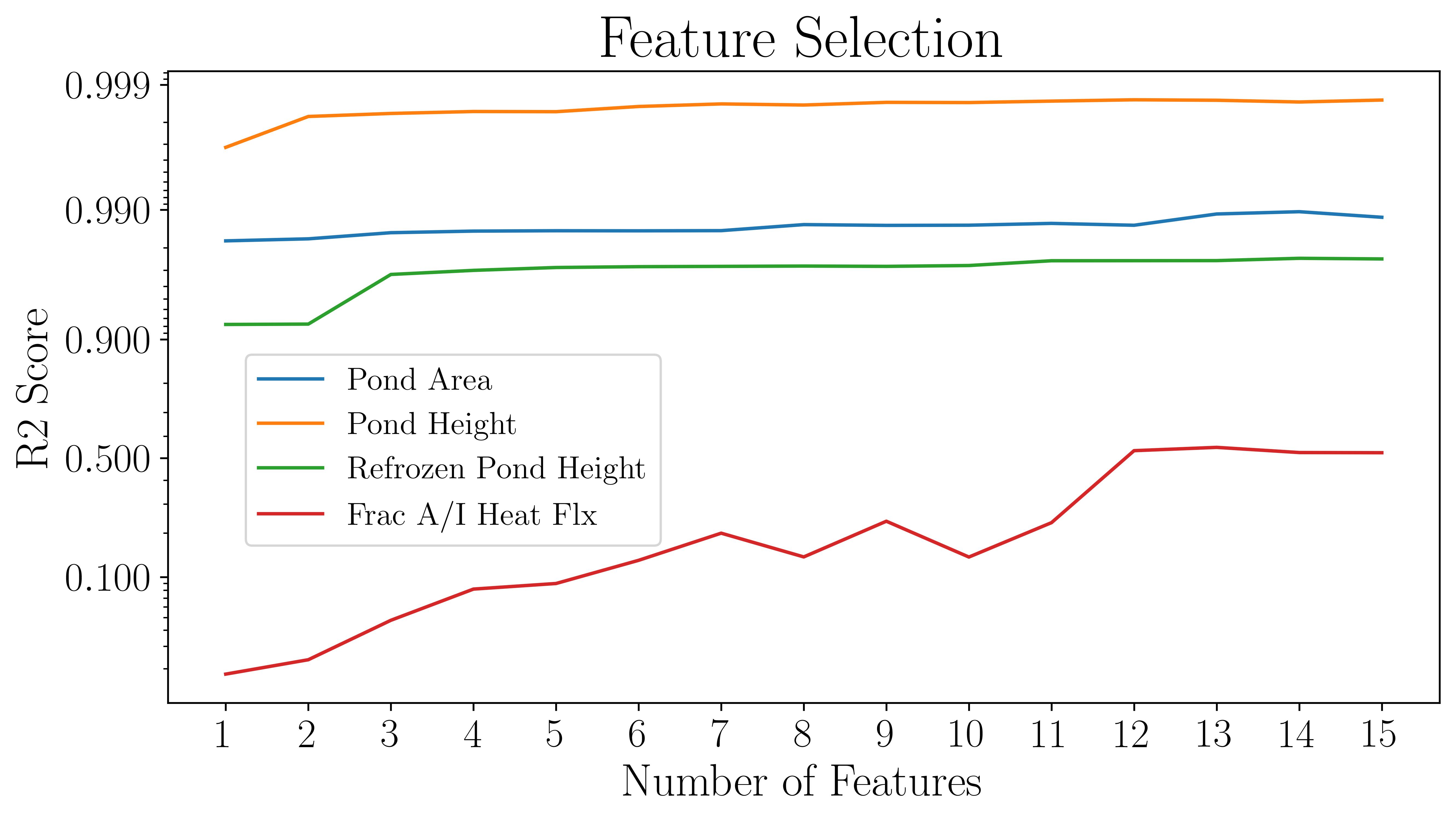}
\caption{R2 scores for each target with respect to the number of features included in training. Neural networks with identical architecture are trained for 25 epochs using $n=1,...,15$ features, for targets pond area, pond height, ipond height and fraction of atmosphere-ice heat flux used to melt frozen lids. The logarithmic scale is used in the $y-$axis.}
 \label{fig:mi_feature_selection_r2_mse_25_epochs}
\end{figure}

Identical NNs, as used in Icepack-NN (Fig.~\ref{fig:nn_architecture}), are trained using $n=1,...,15$ features (see Tab.~\ref{tab:icepack_output_mpp}) ranked by mutual information scores for each target. For example, for pond area, based on the results in Fig.~\ref{fig:mi_feature_selection}, this implies training NNs as follows: for $n=1$ using pond area at previous timestep as the sole feature, for $n=2$, pond area and pond height at previous timestep, for $n=3$ pond area, pond height at previous timestep together with surface temperature, and so on, until all features are included ($n=15$). This is repeated for all four target variables, making a total of $60$ NNs trained for $25$ epochs each. The R2 scores of these NNs in the offline emulation setting are shown in Fig.~\ref{fig:mi_feature_selection_r2_mse_25_epochs}.

Consistent with mutual information scores, we observe that for pond area, height and frozen lid height very good R2 scores are attained with as few as the (appropriate) one, two or three inputs. Recall from Fig.~\ref{fig:mi_feature_selection} that these very influential features are the same variables at the previous timestep. This suggests that one may produce a good emulator of melt ponds knowing their values at previous timestep and a few key variables such as surface temperature. However, for the fraction of atmosphere/ice heat flux (red line), many more factors concur to determining its value. As a consequence, R2 scores of NNs with few features ($<8$) are very low, and effectively the full set of features is needed to reach R2 scores as good as those seen for the first three variables with only few ($\leq 3$) features.  

\begin{figure}
\centering
\includegraphics[width=0.9\textwidth]{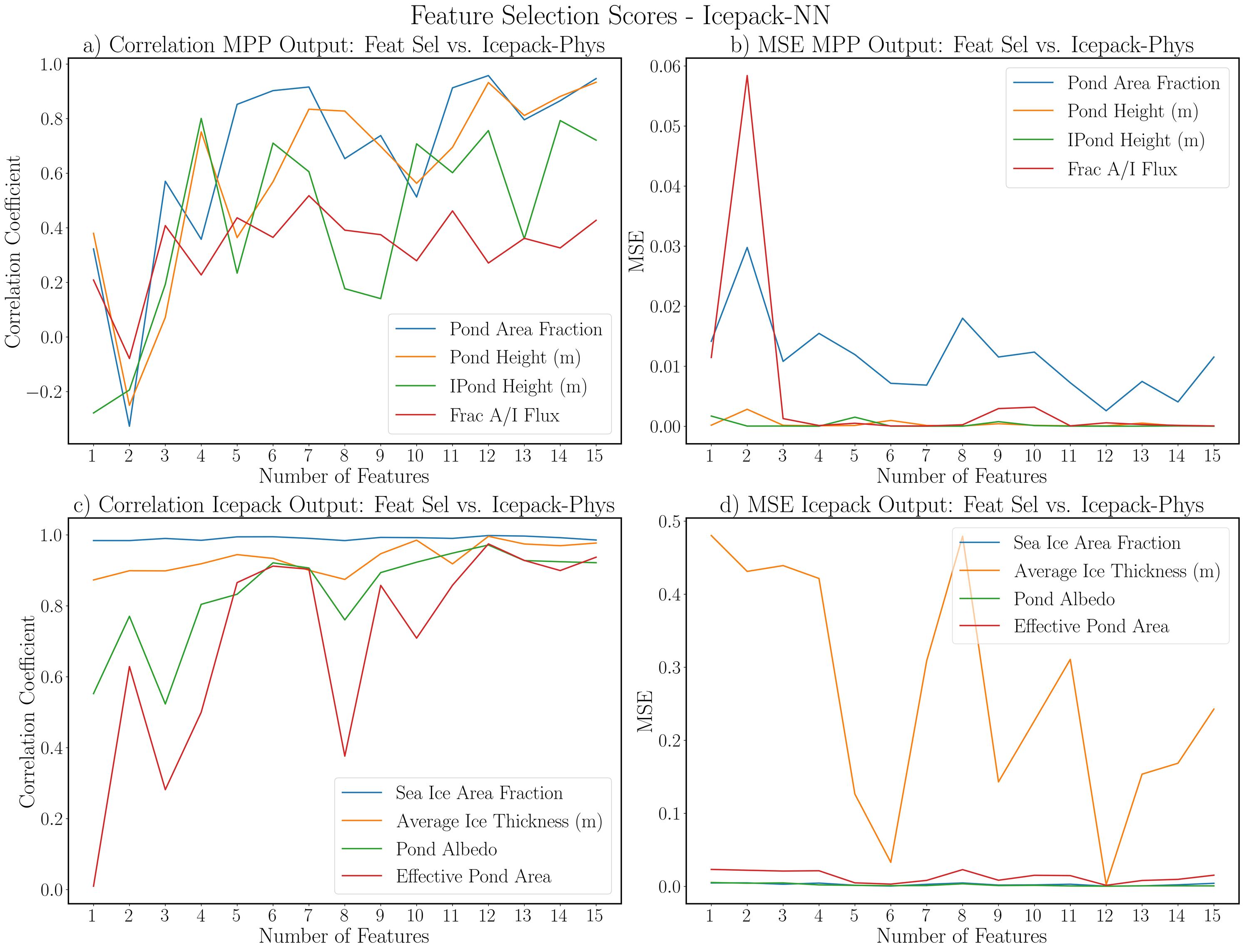}
\caption{Feature selection performance for online simulation. MSE and Correlation Coefficient scores.}
\label{fig:mi_feature_selection_mse_25_epochs_all_feats_online}
\end{figure}

We repeat now the same inspection on the needed features in the context of online emulation and, for $1$ to $15$ features we run Icepack-NN over all 20 test locations (i.e. Icepack-NN where four separate NNs use 1 different feature, 2 features, 3 features, and so on, in order of importance as given by the mutual information criterion). Note that the case $n=15$ coincides with the Icepack-NN with full features discussed in Sect.~\ref{ML-MPP_online}. Figure~\ref{fig:mi_feature_selection_mse_25_epochs_all_feats_online} shows the mean squared error (MSE; panels a and c) and Pearson correlation values (panels b and d) for the four melt pond target variables. Whilst two to three features have shown to suffice for returning offline results that are similar to the MPP, when run online this is not sufficient to replicate the performance of Icepack-Phys; $4$ to $6$ variables would be necessary to emulate and substitute a physical MPP. 

\section{Discussion and future work} \label{disc_and_future}

 By modifying the sea ice albedo, melt ponds play a central role in regulating the energy balance in the Arctic region \citep{Polashenski2012}. Melt pond parametrizations (MPPs) have thus became crucial in present-day sea ice models \citep{Flocco2010, Flocco2012}. Yet the multi scale processes leading the formation of a melt pond are not fully understood, nor they can all be explicitly resolved in sea ice models. Nowadays, and likewise most of the parametrizations in climate models, MPPs are derived from physical principles and have the virtue to also generally respect physical and dynamical balances. Nevertheless, they are also often over parametrized, i.e. a large number of uncertain parameters have to be (possibly simultaneously) tuned. In this context, a further complication arises if the model exhibits high sensitivity to the parameters in the MPPs, whose accurate determination is then paramount, although extremely difficult. 

 We have studied a known advanced MPPs, the level-ice MPP \citep{Hunke2013}, included in a prototypical 1d state-of-the-art model of the sea-ice thermodynamics, Icepack \citep{Hunke2021} We performed a quantitative examination of the its sensitivity to the MPP parameters using the Sobol Sensitivity Analysis \citep{Sobol2001}. Results indicated that the sensitivities of Icepack to its MPP is very large and inhomogeneous. Parameters significantly impact Icepack predicted output both in time and space. As a consequence, even little differences in the MPP parameter values lead to substantially different Icepack behaviours. The Icepack response is not only different in the long term but it also display a strong geographical dependency. The unavoidable uncertainty in the MPP parameters, together with the MPP high sensitivity to parameters themselves, can be the source of large sea-ice prediction errors. 

In the second part of this study we have thus investigated whether machine learning (ML) and in particular neural networks (NNs) are (i) able to learn the physical processes in the physical MPPs, and, (ii) do that with a smaller amount of inputs than those needed by the original physical MPPs. Neural networks were shown to be capable of learning and emulating the MPP. In particular, although simpler linear regression models shown capabilities of learning the input-to-output processes in the MPP (offline emulation) with some skill, only nonlinear NNs were then able to successfully substitute the original physical MPP in the full Icepack (online emulation). In contrast, with the MPP emulator based on linear regression, Icepack fails in reproducing the original Icepack with physical MPP.   

Our results indicate that NNs can learn and represent a viable subgrid-scale parametrisation, and also that a reduced, simpler data driven emulator can act as a MPP in the Icepack model. Feature selection and the associated trained NNs demonstrate that the melt pond parametrisation is too complex for successful online emulation with only one or two features, but is successfull with fewer than all of the input variables requested in the physical MPP. We showed that performance improves when some melt pond properties (pond area fraction, pond height and refrozen lid height) at the previous timestep are known, along with a few other fundamental variables, such as near surface and air temperatures and snow melt rate. In summary, our results suggest that key thermodynamic variables such as temperature, conservation of mass related variables regarding available meltwater (e.g. melt rate of snow), and knowledge of the melt pond at the previous timestep may be key when focusing on an observationally trained emulator.


When seeking to emulate melt ponds and replace existing parametrisations, a final goal is not to merely seek to replace the parametrisation, but to improve upon it. This study is a first step toward the design of fully data-driven MPPs. Our results are encouraging and and demonstrate, as others have for other parametrization in climate models (e.g. for clouds see \citet{Krasnopolsky2013a}), that emulators can replace MPPs in regional and global climate models. Our next steps are to build an emulator from observations and use this to substitute the parametrisation in the Icepack model with a data driven parametrisation. A key challenge that we anticipate is the availability of accurate raw data, whereby accuracy is intended here in term of their spatio-temporal distribution, their precision, as well as their direct relation to the physical quantities of interest. We envisage using recently developed combined data assimilation and ML methods that have shown power in handling sparse and noisy data \citep{Cheng2023}. 



\section{Acknowledgements}

The authors acknowledge the support of the project SASIP funded by Schmidt Futures – a philanthropic initiative that seeks to improve societal outcomes
through the development of emerging science and technologies. SD thanks the NERC Earth Observation Data Acquisition and Analysis Service (NEODAAS) for access to compute resources for this study. SD would like to thank Clara Burgard for useful advice regarding setting up the Icepack model. SD would like to thank David Moffat for assistance with using NEODAAS's Massive GPU Cluster for Earth Observation platform. CEREA is a member of Institut Pierre-Simon Laplace. The authors declare no competing interests. 

\bibliography{references_ml_1}

\appendix



\end{document}